\documentclass{aa}
\usepackage[varg]{txfonts}
\usepackage[breaklinks=true]{hyperref}
\hypersetup{
    colorlinks,
    linkcolor={blue},
    citecolor={blue}, 
    urlcolor={blue}
}

\usepackage{natbib,twoopt}
\bibpunct{(}{)}{;}{a}{}{,} 
\makeatletter
\newcommandtwoopt{\citeads}[3][][]{\href{http://adsabs.harvard.edu/abs/#3}%
{\def\hyper@linkstart##1##2{}%
\let\hyper@linkend\@empty\citealp[#1][#2]{#3}}}
\newcommandtwoopt{\citepads}[3][][]{\href{http://adsabs.harvard.edu/abs/#3}%
{\def\hyper@linkstart##1##2{}%
\let\hyper@linkend\@empty\citep[#1][#2]{#3}}}
\newcommandtwoopt{\citetads}[3][][]{\href{http://adsabs.harvard.edu/abs/#3}%
{\def\hyper@linkstart##1##2{}%
\let\hyper@linkend\@empty\citet[#1][#2]{#3}}}
\newcommandtwoopt{\citeyearads}[3][][]%
{\href{http://adsabs.harvard.edu/abs/#3}
{\def\hyper@linkstart##1##2{}%
\let\hyper@linkend\@empty\citeyear[#1][#2]{#3}}}
\makeatother

\graphicspath{{./figures/}}

\begin{document}

\title{Galaxy cluster strong lensing cosmography}
\subtitle{cosmological constraints from a sample of regular galaxy clusters}

\titlerunning{Galaxy cluster strong lensing cosmography}
\authorrunning{G.~B.~Caminha et al.} 

\author{G.~B.~Caminha       \inst{\ref{mpa}}                        
                            \thanks{e-mail address: \href{caminha@mpa-garching.mpg.de}{caminha@mpa-garching.mpg.de}.} \and
        S.~H.~Suyu           \inst{\ref{mpa},\ref{tum},\ref{sinica}}      \and                            
        C.~Grillo           \inst{\ref{unimilano}}         \and
        P.~Rosati           \inst{\ref{unife},\ref{inafbologna}}          
        }
\institute{
Max-Planck-Institut f\"ur Astrophysik, Karl-Schwarzschild-Str. 1, D-85748 Garching, Germany \label{mpa} \and
Technische Universit\"at M\"unchen, Physik-Department, James-Franck Str. 1, 85741 Garching, Germany \label{tum} \and
Institute of Astronomy and Astrophysics, Academia Sinica, 11F of ASMAB, No.1, Section 4, Roosevelt Road, Taipei 10617, Taiwan \label{sinica} \and
Dipartimento di Fisica, Universit\`a  degli Studi di Milano, via Celoria 16, I-20133 Milano, Italy\label{unimilano} \and
Dipartimento di Fisica e Scienze della Terra, Universit\`a degli Studi di Ferrara, Via Saragat 1, I-44122 Ferrara, Italy\label{unife}\and
INAF - Osservatorio Astronomico di Bologna, via Gobetti 93/3, 40129 Bologna, Italy\label{inafbologna}
}

\abstract{
Cluster strong lensing cosmography is a promising probe of the background geometry of the Universe and several studies have emerged, thanks to the increased quality of observations using space and ground-based telescopes.
For the first time, we use a sample of five cluster strong lenses to measure the values of cosmological parameters and combine them with those from classical probes.
In order to assess the degeneracies and the effectiveness of strong-lensing cosmography in constraining the background geometry of the Universe, we adopt four cosmological scenarios.
We find good constraining power on the total matter density of the Universe ($\Omega_{\rm m}$) and the equation of state of the dark energy parameter $w$.
For a flat $w$CDM cosmology, we find $\Omega_{\rm m} = 0.30_{-0.11}^{+0.09}$ and $w=-1.12_{-0.32}^{+0.17}$ from strong lensing only.
Interestingly, we show that the constraints from the Cosmic Microwave Background (CMB) are improved by factors of 2.5 and 4.0 on $\Omega_{\rm m}$ and $w$, respectively, when combined with our posterior distributions in this cosmological model.
In a scenario where the equation of state of dark energy evolves with redshift, the strong lensing constraints are compatible with a cosmological constant (i.e. $w=-1$).
In a curved cosmology, our strong lensing analyses can accommodate a large range of values for the curvature of the Universe of $\Omega_{\rm k}=0.28_{-0.21}^{+0.16}$.
In all cosmological scenarios, we show that our strong lensing constraints are complementary and in good agreement with measurements from the CMB, baryon acoustic oscillations and Type Ia supernovae.
Our results show that cluster strong lensing cosmography is a potentially powerful probe to be included in the cosmological analyses of future surveys.
}


\maketitle

\section{Introduction}
\label{sec:introduction}

\begin{table}[!]
\centering
\small
\caption{Summary of the different strong lensing models.}
\begin{tabular}{l c c c c c c }

\multicolumn{5}{l}{ Cluster RX~J2129} \\
\multicolumn{5}{l}{ $z_{\rm cluster}=0.234$, $\rm N_{src}= 7$, $z_{\rm src}=[0.68 - 3.43]$} \\
\hline
Model ID &  DOF & $\rm N_{free}$ & rms[\arcsec] & $\rm \chi^2/DOF$ \\
\hline
R2129 fixed                           & 22 & 8 & 0.20 & 0.15 \\ 
R2129 $\Omega_{\rm m}$                & 21 & 9 & 0.19 & 0.15 \\ 
R2129 $\Omega_{\rm m},w$              & 20 & 10& 0.19 & 0.16 \\ 
R2129 $\Omega_{\rm m},\Omega_{\rm k}$ & 20 & 10& 0.19 & 0.15 \\ 
R2129 $\Omega_{\rm m},w_0, w_{\rm a}$ & 19 & 11& 0.19 & 0.16 \\ 
\hline
\\
\multicolumn{5}{l}{ Cluster Abell~S1063} \\
\multicolumn{5}{l}{ $z_{\rm cluster}=0.348$, $\rm N_{src}=20$, $z_{\rm src}=[0.73 - 6.11]$} \\
\hline
Model ID &  DOF & $\rm N_{free}$ & rms[\arcsec] & $\rm \chi^2/DOF$ \\
\hline
A1063 fixed                           & 56 & 14& 0.37 & 0.53 \\ 
A1063 $\Omega_{\rm m}$                & 55 & 15& 0.36 & 0.53 \\ 
A1063 $\Omega_{\rm m},w$              & 54 & 16& 0.36 & 0.54 \\ 
A1063 $\Omega_{\rm m},\Omega_{\rm k}$ & 54 & 16& 0.36 & 0.54 \\ 
A1063 $\Omega_{\rm m},w_0, w_{\rm a}$ & 53 & 17& 0.36 & 0.54 \\ 
\hline
\\
\multicolumn{5}{l}{ MACS~J1931} \\
\multicolumn{5}{l}{ $z_{\rm cluster}=0.352$, $\rm N_{src}=7$, $z_{\rm src}=[1.18 - 5.34]$} \\
\hline
Model ID &  DOF & $\rm N_{free}$ & rms[\arcsec] & $\rm \chi^2/DOF$ \\
\hline
M1931 fixed                           & 12 & 12& 0.38 & 0.91 \\ 
M1931 $\Omega_{\rm m}$                & 11 & 13& 0.37 & 0.96 \\ 
M1931 $\Omega_{\rm m},w$              & 10 & 14& 0.37 & 1.05 \\ 
M1931 $\Omega_{\rm m},\Omega_{\rm k}$ & 10 & 14& 0.38 & 1.07 \\ 
M1931 $\Omega_{\rm m},w_0, w_{\rm a}$ & 9  & 15& 0.35 & 1.06 \\ 
\hline
\\
\multicolumn{5}{l}{ MACS~J0329} \\
\multicolumn{5}{l}{ $z_{\rm cluster}=0.450$, $\rm N_{src}=9$, $z_{\rm src}=[1.31 - 6.17]$} \\
\hline
Model ID &  DOF & $\rm N_{free}$ & rms[\arcsec] & $\rm \chi^2/DOF$ \\
\hline
M0329 fixed                           & 12 & 16& 0.24 & 0.43 \\ 
M0329 $\Omega_{\rm m}$                & 11 & 17& 0.23 & 0.46 \\ 
M0329 $\Omega_{\rm m},w$              & 10 & 18& 0.21 & 0.40 \\ 
M0329 $\Omega_{\rm m},\Omega_{\rm k}$ & 10 & 18& 0.23 & 0.50 \\ 
M0329 $\Omega_{\rm m},w_0, w_{\rm a}$ & 9  & 19& 0.21 & 0.44 \\ 
\hline
\\
\multicolumn{5}{l}{ MACS~J2129} \\
\multicolumn{5}{l}{ $z_{\rm cluster}=0.587$, $\rm N_{src}=11$, $z_{\rm src}=[1.05 - 6.85]$} \\
\hline
Model ID &  DOF & $\rm N_{free}$ & rms[\arcsec] & $\rm \chi^2/DOF$ \\
\hline
M2129 fixed                           & 40 & 14& 0.56 & 1.18 \\ 
M2129 $\Omega_{\rm m}$                & 39 & 15& 0.56 & 1.20 \\ 
M2129 $\Omega_{\rm m},w$              & 38 & 16& 0.55 & 1.19 \\ 
M2129 $\Omega_{\rm m},\Omega_{\rm k}$ & 38 & 16& 0.55 & 1.23 \\ 
M2129 $\Omega_{\rm m},w_0, w_{\rm a}$ & 37 & 17& 0.55 & 1.31 \\ 
\hline
\end{tabular}
\label{tab:cosmo_models}
\tablefoot{The columns are the cluster model ID (including the cluster name and the cosmological model), the degrees of freedom (DOF), the number of free parameters ($\rm N_{\rm free}$), the root mean square (rms) of the differences between the observed and model predicted image positions, and the image position $\chi^2$ per DOF (i.e., the reduced $\chi^2$) for positional uncertainties of 0\farcs5.  For the cosmological model specified in the first column, the listed cosmological parameters indicate the parameters that are allowed to vary. For each cluster we also quote the redshift ($z_{\rm cluster}$), the number of multiply lensed sources ($\rm N_{\rm src}$) and their redshift range ($z_{\rm src}$). The models with fixed cosmology assume a flat-$\Lambda$CDM scenario with $\Omega_{\rm m}=0.3$.}
\end{table}

Cosmological observations suggest that the Universe is mostly composed of somewhat unconventional components whose nature is still not fully understood. 
Observations of the cosmic microwave background \citep[CMB,][]{1992ApJ...396L...1S,2013ApJS..208...19H,2020A&A...641A...6P}, baryon acoustic oscillations \citep[BAO,][]{2002MNRAS.330L..29E,2005ApJ...633..560E,2021arXiv210513549D} and Type Ia supernovae \citep[SNe,][]{1998AJ....116.1009R,1999ApJ...517..565P,2006A&A...447...31A,2018ApJ...859..101S} indicate that the evolution of the Universe at large scales ($>1$~Mpc) is well described by the concordance $\Lambda$CDM model.
These ``classical'' probes converge to a scenario where baryonic and cold dark matter (CDM) accounts for $\approx 30\%$ of the energy density of the Universe.
The remaining $\approx 70\%$ is composed of dark energy associated with the cosmological constant $\Lambda$, and the Universe geometry must be very close to flat, i.e. with vanishing curvature ($\Omega_{\rm k} \approx 0$).

However, deviations from these values and different models describing the dark energy can be accommodated by the current data, see e.g. \citet{2021Univ....7..163M} for a recent review.
Moreover, at smaller scales, the flat {$\Lambda$}CDM model has difficulties to explain some properties related to structure formation, for instance the sub-halo population in galaxy clusters \citep[][]{2015ApJ...800...38G, 2020ApJ...902..124C, 2020Sci...369.1347M} and the inner slope of dark matter halos \citep[][]{2004ApJ...604...88S, 2004ApJ...616...16G, 2011ApJ...728L..39N, 2013ApJ...765...24N, 2013ApJ...765...25N, 2011arXiv1108.5736G, 2012MNRAS.422.3081M, 2015MNRAS.452..343S}, when comparing simulations with observations.
These issues motivate further tests on the $\Lambda$CDM model, both at small and large scales, and play an essential role in the concept and design of cosmological observations and projects, such as the  Baryon Oscillation Spectroscopic Survey \citep{2013AJ....145...10D}, the Dark Energy Survey \citep{2018PhRvD..98d3526A, 2021arXiv210513549D, 2021arXiv210513546P}, the Kilo-Degree Survey \citep{2018MNRAS.476.4662V, 2018MNRAS.474.4894J} and the Planck satellite \citep{2020A&A...641A...6P}, to mention a few.

Strong gravitational lensing, among many other applications in astrophysics (see e.g. \citealt{1992grle.book.....S} and \citealt{2011A&ARv..19...47K}), can also be used to probe the background geometry of the Universe.
On galaxy scales, significant progress has been made in the last decade in order to increase the accuracy of the lens modelling and thus to obtain precise measurements of some relevant cosmological parameters \citep[see e.g.][]{2010ApJ...711..201S, 2019MNRAS.490.1743C, 2020MNRAS.498.1420W, 2020MNRAS.494.6072S, 2020A&A...643A.165B}.
In these systems, the primary sensitivity is on the Hubble constant from the measurements of time delays between multiple images of the same source, especially of strongly lensed quasars \citep[e.g.,][]{2002ApJ...581..823F, 2018A&A...609A..71C, 2020A&A...642A.193M, 2020A&A...640A.105M}.
Although measurements of the values of $\Omega_{\rm m}$ and $\Omega_{\rm k}$ are also possible when the lens galaxy dynamical data is available \citep{2008A&A...477..397G, 2012JCAP...03..016C} or when two or more sources at different redshifts are multiply lensed \citep{2014MNRAS.443..969C, 2021MNRAS.505.2136S},
they are either observationally expensive, rare and are prone to intrinsic mass-sheet degeneracies because of the small number of multiply lensed sources \citep{2014A&A...568L...2S}.

In contrast, massive galaxy clusters can generate dozens of multiple images from sources in a large interval of redshifts.
Because of that, they are excellent systems to probe $\Omega_{\rm m}$, $\Omega_{\rm k}$ and also the equation of state parameter $w$ of the dark energy.
Cluster strong lensing cosmography has been discussed in the past \citep{1992ARA&A..30..311B}, however it requires high-quality data, more specifically, high-resolution imaging and deep spectroscopy, to identify and measure redshifts of a large number of multiple images. 
The possibility of constraining cosmological parameters with this methodology was explored in more detail in subsequent works using simulated data \citep{1998ApJ...502...63L, 2002A&A...387..788G,2009MNRAS.396..354G,2011MNRAS.411.1628D}.
It was only with the combination of imaging with the \emph{Hubble} Space Telescope (HST) and extensive spectroscopy from different instruments in the field of the galaxy cluster Abell~1689 that the first competitive cosmological constraints were obtained in \citet{2010Sci...329..924J}.
Moreover, a collection of models of dark energy with different equations of state was studied using the constraints from Abell~1689 \citep{2015ApJ...813...69M, 2018ApJ...865..122M}.
Later, with the deployment of the Multi Unit Spectroscopic Explorer \citep[MUSE;][]{2014Msngr.157...13B} at the Very Large Telescope (VLT), the number of spectroscopically confirmed multiple images in clusters has increased significantly \citep[see e.g.][]{2015MNRAS.446L..16R, 2016ApJ...822...78G, 2017A&A...600A..90C}.
Based on MUSE spectroscopy and HST data, in \citet{2016A&A...587A..80C} we have shown that the combination of a regular total mass distribution and a large number of multiple images with redshifts in the range of $z_{\rm src}=[0.7-6.1]$ makes the cluster Abell~S1063 an excellent system for cluster strong lensing cosmography. 
Finally, in \citet{2018ApJ...860...94G,2020ApJ...898...87G}, a similarly high-precision lens model of MACS~J1149, one of the very few galaxy clusters with measured time delays between the multiple images of a time-varying source, was used to measure the values of the Hubble constant, $\Omega_{\rm m}$ and $\Omega_\Lambda$.

Although the number of galaxy clusters with accurate strong lensing models has increased in the last years, a combined cosmographical analysis using a cluster sample has not been carried out yet.
In addition to improving the figure of merit, the combination of different clusters is important to mitigate possible systematic effects that might affect the individual strong-lensing models in different ways, for instance, line-of-sight perturbers, intrinsic degeneracies in the models that depend on the cluster redshift and distribution of background sources, mass components external to the cluster cores, etc.
In this work, we use a sample of five clusters, acting as strong lenses, with deep spectroscopy from MUSE and HST photometry from the Cluster Lensing And Supernova survey with Hubble \citep[CLASH;][]{2012ApJS..199...25P} with well studied lens models in the literature, including Abell~S1063 that also belongs to the Hubble Frontier Fields \citep[HFF;][]{2017ApJ...837...97L}.

This paper is structured as follows.
In Section \ref{sec:cluster_sample_and_strong_lensing_models} we present our cluster sample and some aspects of the strong lens modelling.
The methodology and the cosmological constraints from the lensing analyses are shown in Section \ref{sec:cosmological_models}, and the combination with other probes in Section \ref{sec:combination_with_other_probes}. Finally, in Section \ref{sec:conclusions} we summarise our results and discuss future developments of this work.

\section{Cluster sample and strong lensing models}
\label{sec:cluster_sample_and_strong_lensing_models}

\begin{figure}
      \includegraphics[width = 1\columnwidth]{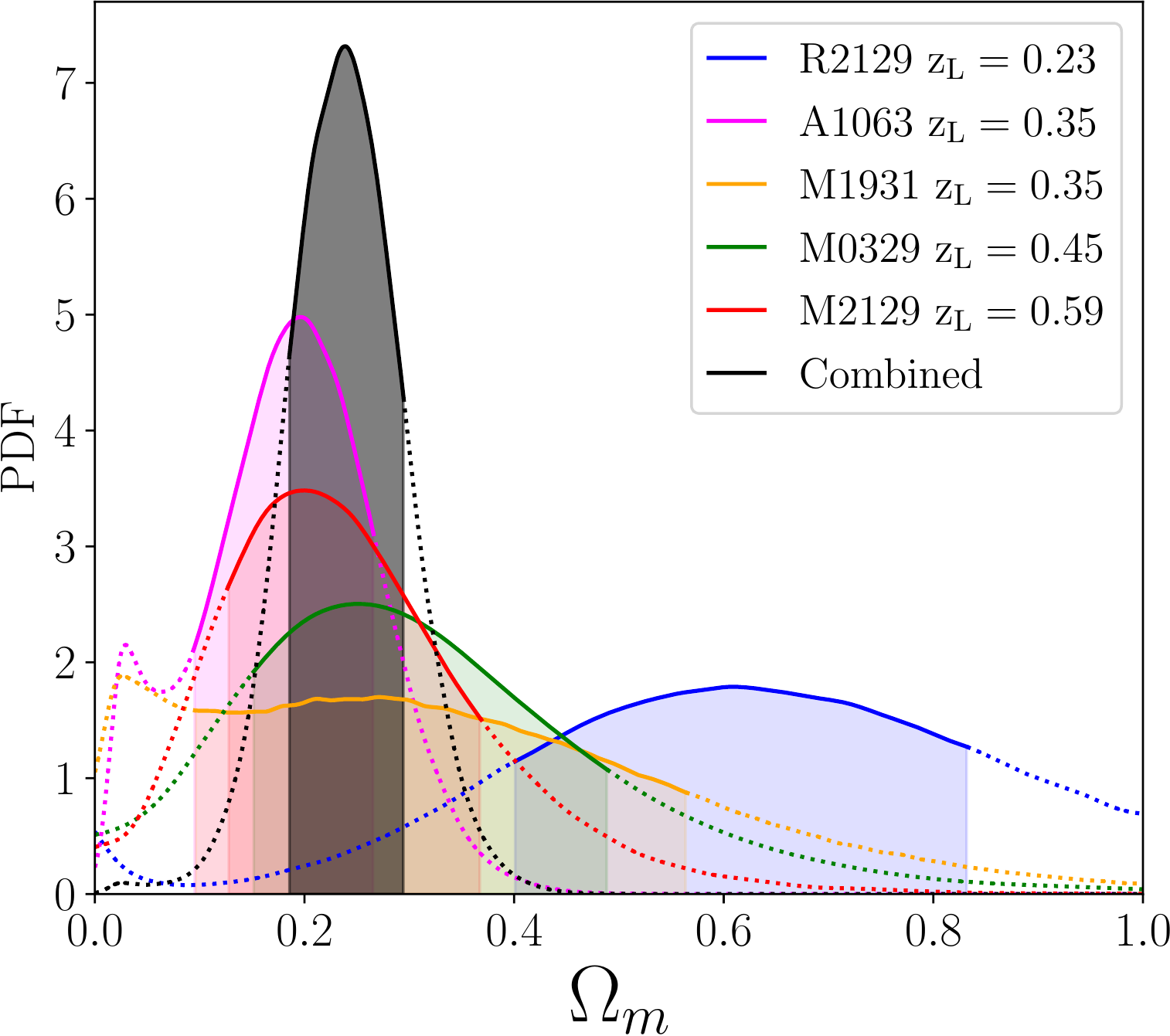}
  \caption{One-dimensional PDFs for the flat-$\Lambda$CDM cosmology, where only the value of $\Omega_{\rm m}$ varies. Strong lensing constraints of each cluster are shown in different colours. Filled regions indicate the 68\% confidence intervals. The grey curve shows the combined constraints from all five clusters. Values and combination with other cosmological probes are presented in Table \ref{tab:cosmo_params}.}
  \label{fig:base_cosmo_hist}
\end{figure}

\begin{figure*}
      \includegraphics[width = 1\columnwidth]{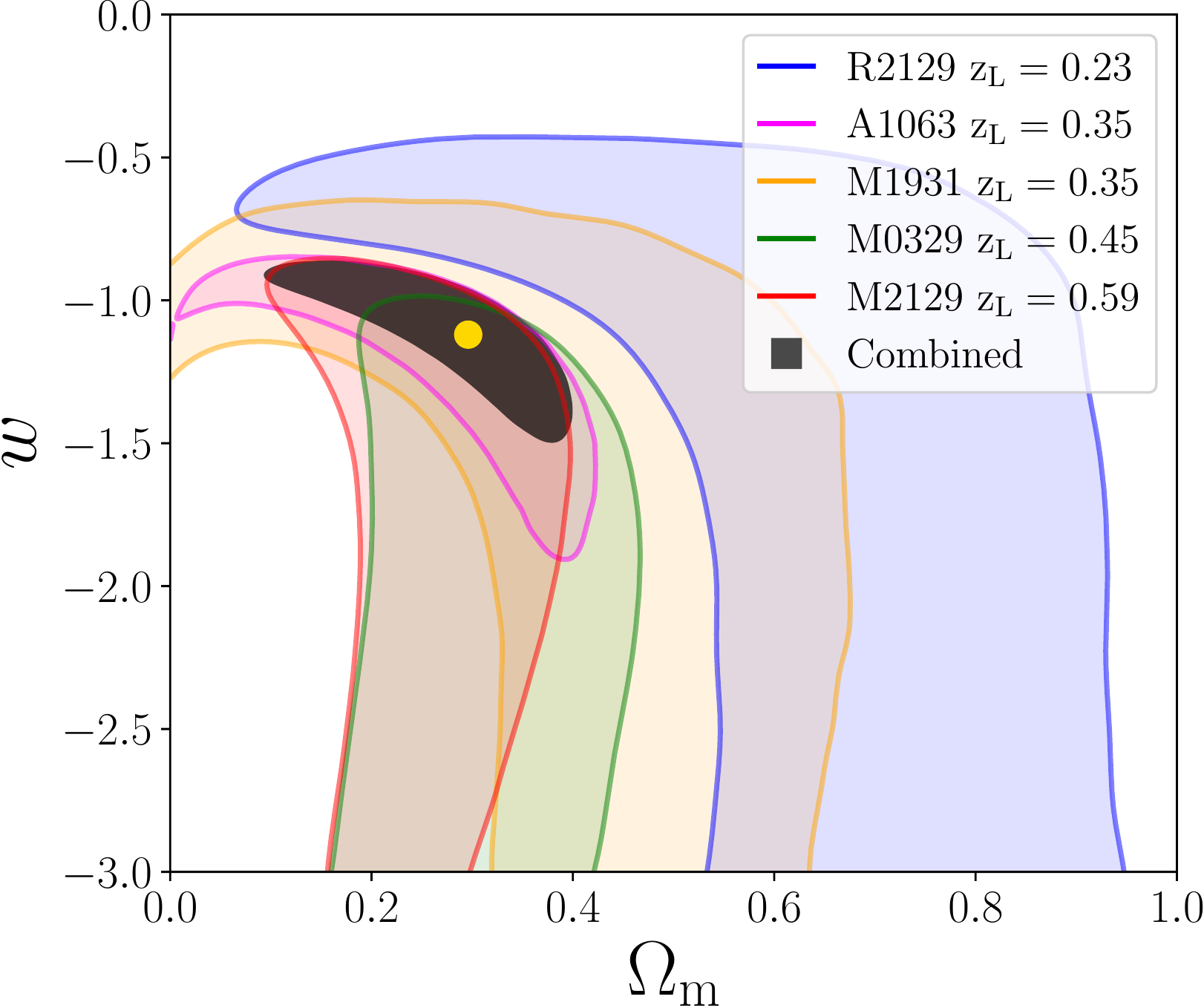}
      \includegraphics[width = 1\columnwidth]{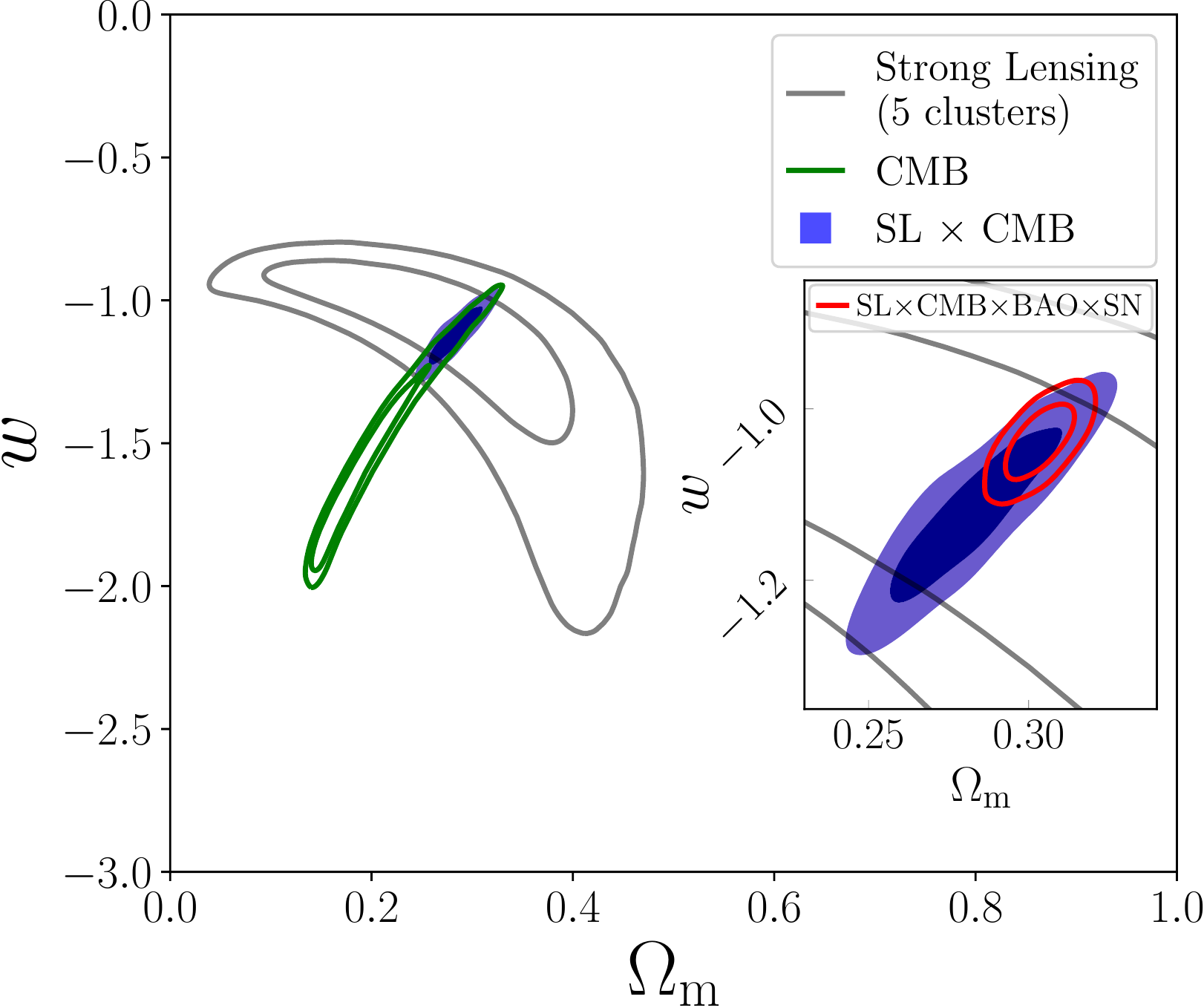}
  \caption{Cosmological constraints for the flat-$w$CDM cosmological model, where $\Omega_{\rm m}$ and $w$ are free parameters. Left panel: 68\% confidence levels from the strong lensing analyses for each cluster (coloured regions) and the combined constraint shown by the grey region. The yellow circle shows the median value of the constrained cosmological parameters (see Table \ref{tab:cosmo_params}). Right panel: combination with other cosmological observables. The inset box shows the additional combination with BAO and SN. The contours indicate the 68\% and 95\% confidence regions. The strong lensing confidence levels are almost perpendicular to those obtained from the CMB and when combined, strong lensing improves the CMB constraints by factors of 2.5 and 4.0 on $\Omega_{\rm m}$ and $w$, respectively.}
  \label{fig:flat_cosmo}
\end{figure*}

In this work, we use clusters with strong lensing models based on extensive HST photometry and deep spectroscopy, mainly from VLT/MUSE.
They are the CLASH ``gold sample'' presented in \citet{2019A&A...632A..36C}, consisting of the clusters RX~J2129, MACS~J1931, MACS~J0329 and MACS~J2129.
We have also included the HFF target Abell~S1063, which has been shown to be a very efficient cluster for cosmography, given its combination of regular mass distribution and the high number of strong lensing constraints \citep[see e.g.][]{2016A&A...587A..80C}.
This sample of clusters has a redshift range of $z_{\rm lens}\approx [0.23 - 0.59]$ with a large number of multiple images.
Thanks to dedicated spectroscopic followups \citep{2013ApJ...772..141B, 2014Msngr.158...48R, 2016A&A...587A..80C, 2016A&A...595A.100C, 2016ApJ...823L..14H, 2017A&A...599A..28K, 2017MNRAS.466.4094M, 2019A&A...632A..36C}, many multiple images have spectroscopic confirmation in the range $z_{\rm src} \approx [0.7 - 6.9]$, thus providing exceptional constraints to the strong lensing models.
As pointed out by \citet{2011MNRAS.411.1628D} and \citet{2017MNRAS.470.1809A}, this large redshift range is particularly important to reduce the biases and intrinsic degeneracies on the cosmological constraints obtained from strong lensing analyses, which we describe in Section~\ref{sec:cosmological_models}.
In our models, we use only families of multiple images with spectroscopic confirmation in order not to include any additional free parameter (i.e. the background source redshift) and to ensure we do not have misidentification of multiple images.

We model the mass distribution of our cluster sample using a superposition of parametric profiles.
In detail, the cluster scale components are parameterised by pseudo-isothermal elliptical mass distributions \citep[PIEMD,][]{1993ApJ...417..450K}.
We have also tested the Navarro-Frenk-White model \citep[NFW,][]{1996ApJ...462..563N, 1997ApJ...490..493N} to describe the cluster scale mass distribution, however these models are less accurate in predicting the multiple image positions when comparing to PIEMD models.
In the Appendix \ref{ap:nfw_models}, we discuss the best fit models using the NFW profile.
The cluster members are modelled by dual pseudo-isothermal mass distributions with vanishing core radius \citep[][]{2007arXiv0710.5636E, 2010A&A...524A..94S}, and follow a constant total mass-to-light ratio to reduce the number of free parameters.
For more details on the observational data and the mass modelling of the cluster sample used in this paper, we refer to the works of \citet{2016A&A...587A..80C, 2019A&A...632A..36C}.

In Table \ref{tab:cosmo_models}, we list the degree of freedom ($\rm DOF\equiv$ number of constraints $-$ number of free parameters) and the number of free parameters ($\rm N_{free}$) for each cluster and the different cosmological models we consider in this work.
Moreover, we also quote the root mean square (rms) of the differences between the model predicted and observed positions of the multiple images, and the reduced image-plane $\chi^2$.
The models with fixed cosmology have from eight to 16 free parameters, a number that is relatively simple to sample with commonly used MCMC codes, for instance the \emph{bayeSys} algorithm \footnote{\url{http://www.inference.org.uk/bayesys/}} implemented in the \emph{lenstool} software that we use \citep{1996ApJ...471..643K,2007NJPh....9..447J,2009MNRAS.395.1319J}.
We note that in some models, the presence of a second dark matter halo is favoured by the best-fitting models.
However, these secondary components are located outside the strong lensing region, i.e. at projected distances from the cluster centre larger than $\gtrsim 200$~kpc, acting as perturbers and are not the main contributors for the formation of multiple images used in this work.

Clusters with more complex total mass distributions, for instance merging systems or cases with prominent perturbers along the line-of-sight at different redshifts, demand an increased number of free parameters to predict the positions of all multiple images.
This is the case of some HFF and CLASH clusters, such as MACS~J0416, Abell~370, Abell~2744, MACS~J1149 MACS~J1206 \citep{2016ApJ...822...78G, 2017A&A...600A..90C, 2017A&A...607A..93C,2018MNRAS.473..663M, 2019MNRAS.485.3738L, 2021A&A...645A.140B, 2021A&A...646A..83R}.
These clusters need from 21 to $\approx$ 40 free parameters to characterise their mass components, compared to fewer than 20 free parameters for the clusters in our sample indicated in Table \ref{tab:cosmo_models}.
The high number of free parameters makes it challenging to sample the likelihood, especially the cosmological parameters that have a small impact on the multiple image positions.
Moreover, the fact that they are merging systems and we use a combination of relatively simple parametric models might introduce a bias in our measurement of the cosmological parameters.
One possibility is to explore more complex, or semi-parametric models \citep[see e.g.][]{beauchesne2021improving} to describe such complex mass distributions.
In this work, we focus on the sample of clusters with regular mass distribution, and the cosmographical analyses of clusters with a more complex total mass distribution and/or in a merging state will be the subject of a future publication.

\section{Strong lensing cosmological constraints}
\label{sec:cosmological_models}

In this section we describe the methodology and assumptions we adopt in our cluster strong lensing models to probe the background cosmology of the Universe.
The results of our strong lensing analyses, i.e. the constrained parameters and their posterior distributions, are presented at the end of this section.

\subsection{Distance ratios from multiply lensed sources}
\label{sec:cosmological_models:lensdist}

Through its relation to the angular diameter distances, the strong lensing effect is affected by the background geometry of the Universe.
The relation between the observed ($\vec{\theta}$) and intrinsic ($\vec{\beta}$) positions of a lensed galaxy $i$, is given by the so-called lens equation:
\begin{equation}
\vec{\beta}_{i} = \vec{\theta}_{i} - \frac{D_{L,Si}}{D_{O,Si}} \vec{\alpha}(\vec{\theta}_{i}),
\label{eq:lens_equation}
\end{equation}
where $D_{L,Si}$ and $D_{O,Si}$ are the angular diameter distances between the lens ($L$) and the background source ($S_{i}$), and the observer ($O$) and the background source, respectively.
The function $\vec{\alpha}$ is the deflection angle and this quantity depends only on the projected total lens potential $\psi(\vec{\theta})$ of the deflector and the position of the lensed image via the relation $\vec{\alpha}(\vec{\theta}) \propto  \vec{\nabla} \psi\left(\vec{\theta}\right)$.
From this equation, the ratio of the cosmological distances and the lens potential are degenerate through a multiplicative factor.
However, in the case of more than one multiple-image family at different redshifts, this degeneracy is strongly reduced since the projected lens potential of the cluster is the same for all sources.
Thus, the positions of the multiple images of two background sources $S_i$ and $S_j$ provide information on the quantity
\begin{equation}
\Xi_{ij} \equiv \frac{D_{L,Si}D_{O,Sj}}{D_{L,Sj}D_{O,Si}}.
\label{eq:family_ratio}
\end{equation}
For $\rm N_{\rm s}$ background sources that are strongly lensed by the same cluster, we therefore obtain $\rm N_{\rm s}-1$ independent distance ratios $\Xi_{ij}$.
The measurements of these $\Xi_{ij}$ then yield information on the background cosmology.
Specifically in our cluster sample, the number of multiply lensed background sources used as input in the lens models are 7, 20, 7, 9 and 11 for RXJ~2129, Abell~S1063, MACS~J1931, MACS~J0329 and MACS~J2129, respectively.

\subsection{Cosmological models}
\label{sec:cosmological_models:cosmomod}

We consider four different cosmological models to assess the degeneracies and the effectiveness of strong-lensing cosmography in constraining the background geometry of the Universe.  They are:

\begin{itemize}

\item flat-$\Lambda$CDM cosmology is the simplest model adopted here. It is a flat Universe where only $\Omega_{\rm m}$ varies and the other parameters are fixed ($w=-1$, $w_{\rm a}=0$, and $\Omega_{\rm k} = 0$).

\item flat-$w$CDM is also a flat cosmology in which the dark energy equation of state parameter ($w\equiv P/\rho$) is allowed to vary, but it does not depend on redshift (i.e. $w_{\rm a} = 0$).

\item curved-$\Lambda$CDM is a cosmological model with free curvature and fixed equation of state of dark energy ($w=-1$) .

\item CPL cosmology consists of a flat Universe where the dark energy equation of state can vary with redshift using a Chevallier-Polarski-Linder parametrization \citep[CPL,][]{2001IJMPD..10..213C, 2003PhRvL..90i1301L} given by:
\begin{equation}
w(z) =  w_0 + w_{\rm a}\frac{z}{1+z}.
\label{eq:cpl_parametrization}
\end{equation}
\end{itemize}
Moreover, we adopt flat priors for the values of the cosmological parameters, allowing them to vary in the ranges:
\begin{alignat}{3}
\Omega_{\rm m} &= [0, 1],  \\
\Omega_\Lambda &= [0, 1], \\
w \,({\rm or}\,w_0) &= [-3, 0], \\
w_{\rm a} &= [-3, 3].
\end{alignat}
We note that in \emph{lenstool}, the curvature is parameterised by the parameter $\Omega_\Lambda$ and the results here are presented in terms of $\Omega_{\rm k}$, converted with the linear relation $\Omega_{\rm k} = 1 - \Omega_{\rm m} - \Omega_\Lambda$.

\subsection{Lens modelling}
\label{sec:cosmological_models:lensmod}

\begin{figure*}
      \includegraphics[width = 1\columnwidth]{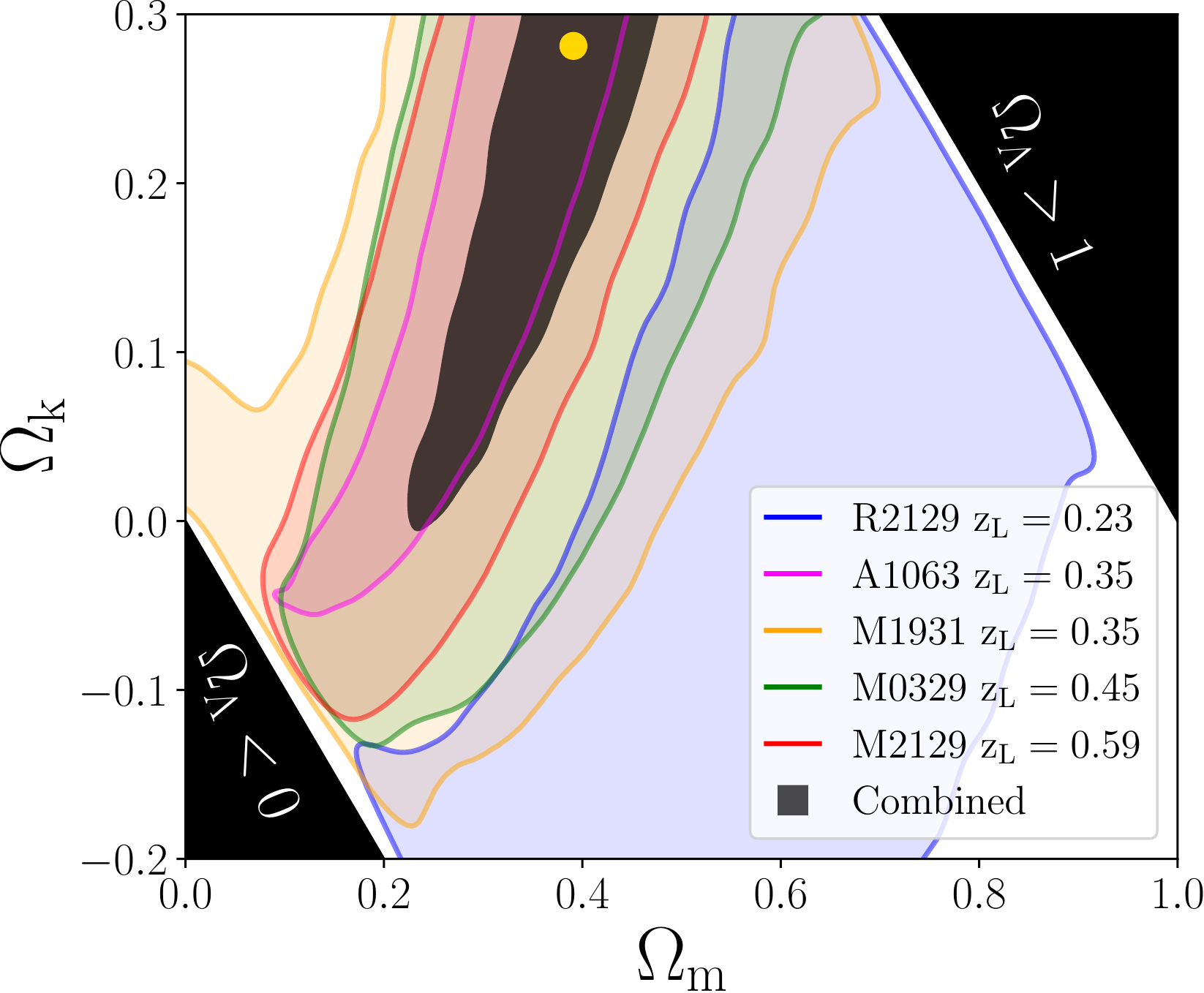}
      \includegraphics[width = 1\columnwidth]{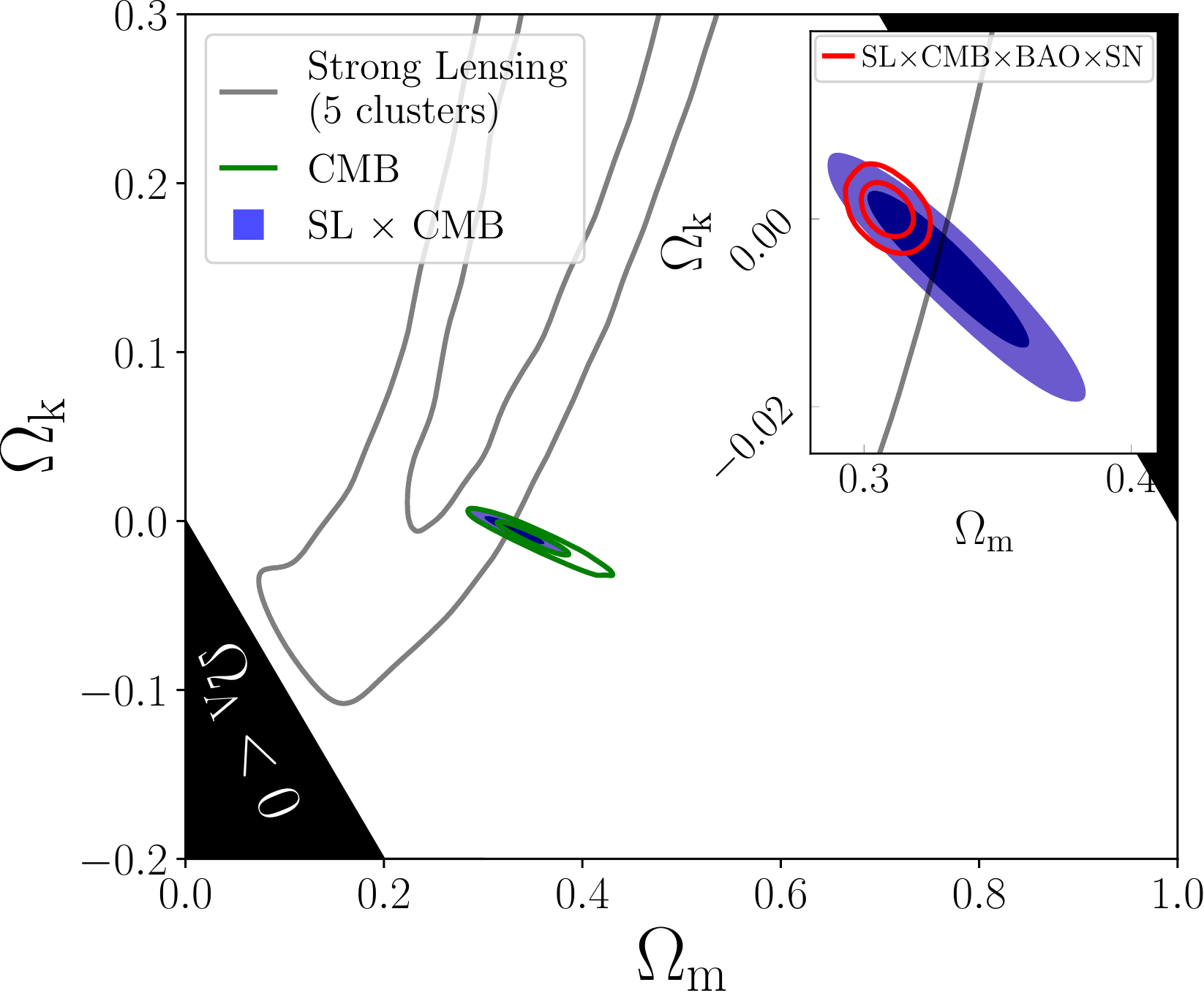}
  \caption{Same as Figure \ref{fig:flat_cosmo}, but for the curved-$\Lambda$CDM cosmology, with fixed $w=-1$ and free $\Omega_{\rm m}$ and $\Omega_{\rm k}$. Regions outside our priors on $\Omega_\Lambda$ (i.e. $\Omega_\Lambda < 0$ or $>1$) are not sampled and indicated in the panels. Although strong lensing only does not obtain stringent constraints on the curvature parameter, it is still complementary to other probes due to its perpendicular parameter degeneracies with respect to the CMB.}
  \label{fig:open_cosmo}
\end{figure*}

In order to obtain the cosmological constraints with our strong lensing models, we first obtain the best-fit parameters for each model (see Table \ref{tab:cosmo_models}) considering a positional error on the multiple image positions of $0\farcs5$, and, in a second step, we run \emph{lenstool} to sample the posterior distribution of the free parameters.
To secure sensible confidence intervals for the values of the parameters (i.e. error bars), we re-scale the positional errors in order to have $\chi^2/DOF = 1$ in the sampling runs.
The values of the re-scaled errors are similar to those of the rms of each model and range from $0\farcs20$ to $0\farcs56$.
With this, we account for effects that are not included in the modelling, such as line-of-sight mass structures \citep{2010Sci...329..924J, 2012MNRAS.420L..18H}, possible deviations of some cluster members from the adopted scaling relations and asymmetries in the cluster scale component that are not well represented by simple elliptical models.
We leave the chains running until they reach at least $10^6$ points for all models and, in the case of the CPL cosmology that has three free cosmological parameters ($\Omega_{\rm m},w_0, w_{\rm a}$), we allow for a longer run to obtain a minimum of $2\times10^6$ points to properly sample the parameter space.
We note that in all models the Gelman-Rubin convergence test \citep{Brooks} results in values lower than 1.1 for all free parameters, which ensures that the chains have reached convergence.

In total, we sample the parameter space of 20 independent models for all five clusters in the four different cosmological models.
From the MCMC chains, we compute the probability distribution functions (PDF) of the cosmological parameters, and the combined constraints are obtained by multiplying the PDFs of each cluster of a specific cosmology, i.e.,
\begin{equation}
P^{\rm total}_{j} = \prod_{i=1}^{5} P_{i}({\rm cosmo}_j),
\end{equation}
where five is the number of clusters we use in this work, $j$ refers to one of the four adopted cosmologies, and $P_{i}$ is the PDF of cluster $i$ marginalised over its total mass parameters.
When combining with other cosmological observables (see Section \ref{sec:combination_with_other_probes}), the function $P^{\rm total}_{j}$ is further multiplied with the PDFs from the additional probes.
To obtain the constrained values of each cosmological parameter, we project the combined PDF in the corresponding direction and compute the median value and confidence interval of the 1-dimensional distribution.

\subsection{Cosmological constraints from lensing clusters}
\label{sec:cosmological_models:result}

In Figure \ref{fig:base_cosmo_hist}, we show the posterior distribution functions (PDFs) of the flat-$\Lambda$CDM cosmological model for each cluster in different colours.
The single PDFs are somewhat broad, with a 68\% confidence level interval of the order of $\approx 0.2-0.4$.
Despite having a relatively small model rms (see Table \ref{tab:cosmo_models}), we note that the PDF of RX~J2129 seems to overestimate the value of $\Omega_{\rm m}$ when comparing to the other clusters, but with a long tail towards low values.
This might be due to the fact that the multiple images are located within a very small region of the cluster core ($<100$~kpc), and their redshift interval is $z_{\rm src} = [0.68 - 3.43]$, whereas the other clusters have at least one multiple image family at $z_{\rm src}>5$.
The more restricted source redshift range in RX~J2129 results in larger uncertainties on $\Omega_{\rm m}$, and the higher $\Omega_{\rm m}$ value is statistically consistent with those of the other clusters within 2$\sigma$.\footnote{With 5 clusters, it is statistically likely that 1 or 2 clusters would yield constraints that are not consistent within 1$\sigma$ with the constraints from other clusters, but are likely to be consistent within 2$\sigma$.}  Therefore, the combined constraint is not significantly affected by this cluster.
From Table \ref{tab:cosmo_models}, the changes on the values of the best-fit rms and reduced $\chi^2$ are very small.
This indicates that the model with fixed cosmology (i.e. $\Omega_{\rm m}=0.3$) is already close to the actual value; thus $\Omega_{\rm m}$ only slightly impacts on the best-fit models.
When combining all clusters, however, we obtain a narrower constraint of $\Omega_{\rm m} = 0.24_{-0.05}^{+0.06}$.

In the left panel of Figure \ref{fig:flat_cosmo}, we show the 68\% confidence regions for the parameters $\Omega_{\rm m}$ and $w$ in the flat-$w$CDM
 scenario (i.e. $\Omega_{\rm k} = 0$).
Similar to Figure \ref{fig:base_cosmo_hist}, the constraints from individual clusters are relatively wide, except for Abell~S1063.
As we mentioned before, and discussed in \citet{2016A&A...587A..80C}, this cluster has a remarkably regular shape in combination with a large number of spectroscopically confirmed multiple images.
Moreover, the intrinsic degeneracies between these two cosmological parameters changes with respect to the lens redshift through the distance ratio constraints in Equation (\ref{eq:family_ratio}).
Lenses at lower redshifts tend to have an extended ``tail'' towards higher values of $\Omega_{\rm m}$ for low values of $w$ (see, e.g., RX~J2129 and MACS~J1931), forming an inverted L-shaped constraint in the $w$-$\Omega_{\rm m}$ plane.
Such L-shaped degeneracy is less pronounced for higher-redshift clusters, such as MACS~J0329 and MACS~J2129.
In this cosmology, the strong lensing combined constraints (median and 68\% confidence levels) are $\Omega_{\rm m}=0.30_{-0.10}^{+0.09}$ and $w=-1.12_{-0.32}^{+0.17}$ (see Table \ref{tab:cosmo_params}).

We also explore the possibility of constraining the curvature of the Universe with our strong lens models.
In order to maintain the number of free parameters describing the curved-$\Lambda$CDM cosmology, we fix the value $w=-1$, and vary those of $\Omega_{\rm m}$ and $\Omega_{\rm k}$.
In the left panel of Figure \ref{fig:open_cosmo}, we show the strong lensing constraints on these two parameters.
Through our prior range, regions of nonphysical values of $\Omega_\Lambda$ are excluded (see Section \ref{sec:cosmological_models:cosmomod}) and indicated in the figure.
In this scenario, the combined constraint on the dark matter density parameter is $\Omega_{\rm m} = 0.39_{-0.11}^{+0.08}$.
The median value is somewhat larger when comparing with the flat-$w$CDM cosmology, but it is consistent within the 68\% confidence interval.
On the other hand, the curvature is not strongly constrained and our models provide a relatively large interval, $\Omega_{\rm k} = 0.28_{-0.21}^{+0.16}$.
Although large, these constraints on $\Omega_{\rm k}$ are still complementary to those from other cosmological probes, as we will discuss in Section \ref{sec:combination_with_other_probes}.

Finally, in Figure \ref{fig:cpl} we show the constraints on the three cosmological parameters of the CPL scenario from our models.
Because of the additional free parameter $w_{\rm a}$ (see Equation (\ref{eq:cpl_parametrization})), the strong lensing constraints of each single cluster are wider and difficult to visualise.
Thus, only the 68\% confidence level of the combined constraints are shown in each 2-dimensional projection of the three parameters.
We note that the region where $w_{\rm a} > -w_0$ is removed because it yields a Universe model dominated by dark energy at early times and this is excluded by high-redshift studies \citep[see e.g.][]{2007ApJ...664..633W,2008ApJ...686..749K, 2011ApJS..192...18K, 2020A&A...641A...6P}.
In this scenario, we obtain the following constraints from the strong lensing only analyses: $\Omega_{\rm m} = 0.35_{-0.11}^{+0.07}$, $w_0 = -1.00_{-0.43}^{+0.32}$ and $w_{\rm a} = -0.95_{-1.31}^{+1.43}$.

In all cosmological scenarios, the values of $\Omega_{\rm m}$ are in agreement within the 68\% confidence levels (see Table \ref{tab:cosmo_params}), although the size of this interval increases with the number of free cosmological parameters, as expected.
Overall, the current precision of the lensing models, which is reached thanks to the deep spectroscopy available, provides meaningful constraints from the strong-lensing-only analyses.
The combination and complementarity of our strong-lensing constraints with other observational probes is discussed in the following section.

\section{Combination with other cosmological probes}
\label{sec:combination_with_other_probes}

\begin{figure}
      \includegraphics[width = 1\columnwidth]{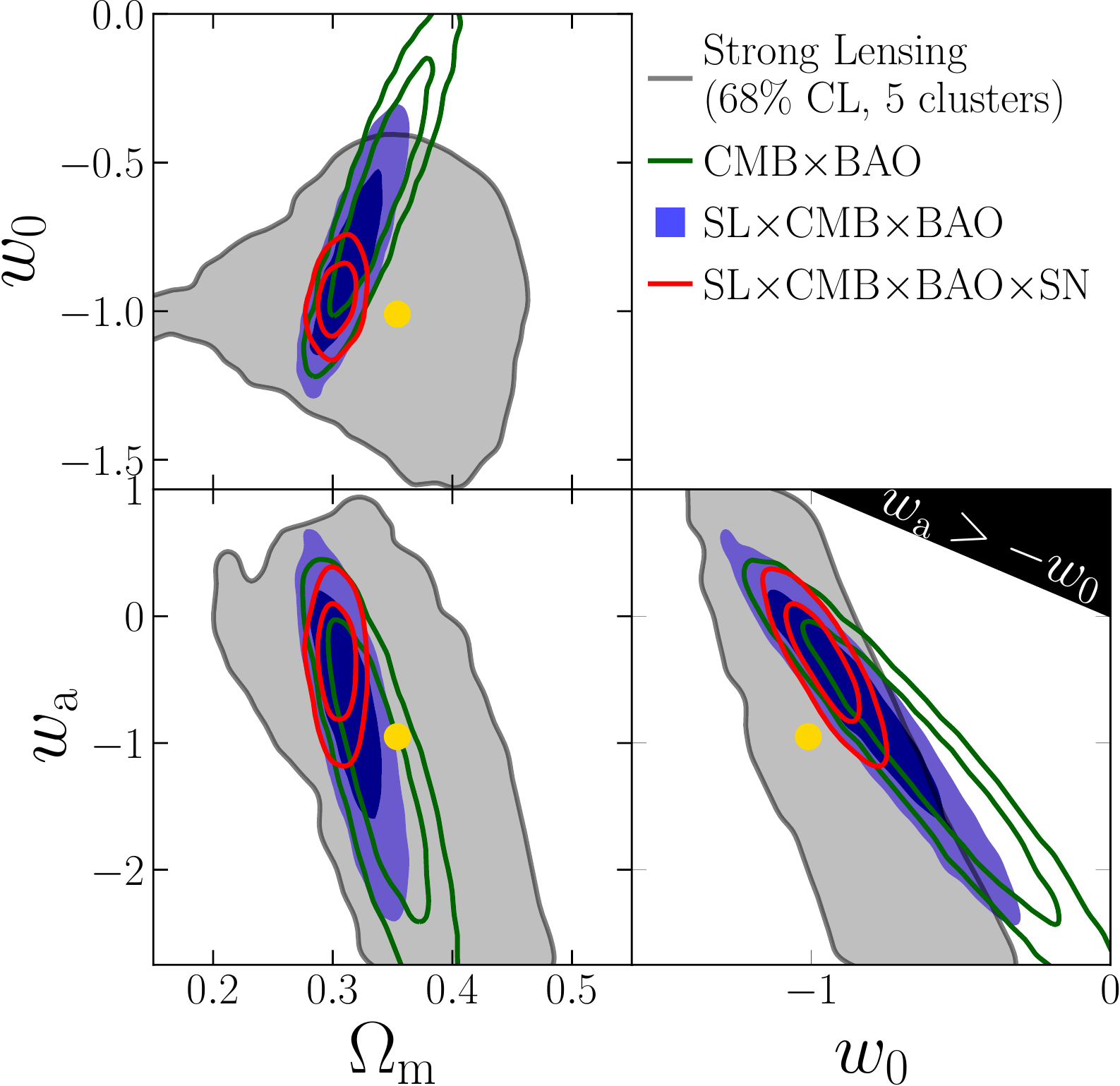}
  \caption{Confidence levels for the CPL model. Here we show in grey the 68\% confidence level for the strong-lensing-only analyses for better visualisation, and the yellow circle indicates the median values of Table \ref{tab:cosmo_params}. The 68\% and 95\% combined constraints of different probes are shown with coloured contours. The region where $w_{\rm a} > -w_0$ is excluded because it produces a non-physical cosmology. In this cosmology with one additional free parameter, our strong lensing models are still capable of improving the constraints when combined to classical probes, especially on the parameters $w_0$ and $w_{\rm a}$ of the dark energy equation of state.}
  \label{fig:cpl}
\end{figure}

\begin{table*}[!]
\renewcommand{\arraystretch}{1.5}

\caption{Summary of the cosmological constraints.}
\begin{tabular}{l c c c  c c c c } \hline \hline
flat-$\Lambda$CDM  & $\Omega_{\rm m}$ & $\delta \Omega_{\rm m}$\\
\hline
SL                                 & $0.239_{-0.054}^{+0.056}$ & ---\\
CMB                                & $0.315_{-0.007}^{+0.007}$ & ---\\
SL$\times$CMB                      & $0.314_{-0.007}^{+0.007}$ & 1.01 \\
CMB$\times$BAO$\times$SN           & $0.310_{-0.005}^{+0.006}$ & ---\\
SL$\times$CMB$\times$BAO$\times$SN & $0.310_{-0.005}^{+0.006}$ & 1.00 \\
\hline\hline

flat-$w$CDM & $\Omega_{\rm m}$ & $\delta\Omega_{\rm m}$& $w$& $\delta w$ \\
\hline
SL                                 & $0.296_{-0.105}^{+0.086}$&  --- & $-1.12_{-0.32}^{+0.17}$& --- \\
CMB                                & $0.186_{-0.032}^{+0.057}$&  --- & $-1.60_{-0.23}^{+0.31}$& --- \\
SL$\times$CMB                      & $0.283_{-0.018}^{+0.018}$& 2.52 & $-1.12_{-0.07}^{+0.07}$& 4.00 \\
CMB$\times$BAO$\times$SN           & $0.306_{-0.008}^{+0.008}$&  --- & $-1.03_{-0.03}^{+0.03}$& --- \\
SL$\times$CMB$\times$BAO$\times$SN & $0.303_{-0.007}^{+0.007}$& 1.07 & $-1.04_{-0.03}^{+0.03}$& 1.07 \\
\hline\hline

curved-$\Lambda$CDM & $\Omega_{\rm m}$& $\delta\Omega_{\rm m}$ & $\Omega_{\rm k}$ & $\delta \Omega_{\rm k}$\\
\hline
SL                                 & $0.391_{-0.109}^{+0.076}$&  ---   & $\;\,\, 0.2813_{-0.2082}^{+0.1588}$& --- \\
CMB                                & $0.352_{-0.024}^{+0.023}$&  ---   & $-0.0106_{-0.0065}^{+0.0068}$& ---\\
SL$\times$CMB                      & $0.330_{-0.020}^{+0.021}$& 1.15   & $-0.0048_{-0.0062}^{+0.0053}$& 1.16 \\
CMB$\times$BAO$\times$SN           & $0.309_{-0.006}^{+0.006}$&  ---   & $\;\,\, 0.0008_{-0.0020}^{+0.0020}$& --- \\
SL$\times$CMB$\times$BAO$\times$SN & $0.308_{-0.006}^{+0.006}$& 1.03   & $\;\,\, 0.0010_{-0.0020}^{+0.0020}$& 1.00 \\
\hline\hline

CPL  & $\Omega_{\rm m}$ & $\delta\Omega_{\rm m}$& $w_0$& $\delta w_{0}$ & $w_{\rm a}$ & $\delta w_{\rm a}$\\
\hline
SL                                 & $0.354_{-0.105}^{+0.070} $& ---    & $-1.01_{-0.43}^{+0.32} $&  ---    & $-0.95_{-1.31}^{+1.43} $& --- \\
CMB$\times$BAO                     & $0.340_{-0.027}^{+0.027} $& ---    & $-0.59_{-0.28}^{+0.27} $&  ---    & $-1.22_{-0.78}^{+0.75} $& --- \\
SL$\times$CMB$\times$BAO           & $0.315_{-0.019}^{+0.019} $& 1.40   & $-0.84_{-0.19}^{+0.21} $&  1.36   & $-0.71_{-0.65}^{+0.56} $& 1.26 \\
CMB$\times$BAO$\times$SN           & $0.306_{-0.011}^{+0.011} $& ---    & $-0.96_{-0.08}^{+0.09} $&  ---    & $-0.27_{-0.33}^{+0.29} $& --- \\
SL$\times$CMB$\times$BAO$\times$SN & $0.304_{-0.010}^{+0.011} $& 1.03   & $-0.96_{-0.08}^{+0.09} $&  1.02   & $-0.35_{-0.33}^{+0.28} $& 1.01 \\
\hline
\end{tabular}
\label{tab:cosmo_params}
\end{table*}

To explore the full potential of our cluster-strong-lensing cosmological constraints, we combine our results with those from other probes in this section.
In order to be consistent with previous works and quantify the improvement on the figure of merit of the combined constraints, we use the publicly available posterior distributions from the Planck collaboration \citep{2020A&A...641A...6P}.
Specifically, we consider the chains containing constraints from CMB plus lensing power spectrum reconstruction likelihood, since for the CPL cosmology, the CMB-only chain is not available.
The other measurements come from Type Ia Supernovae (SNe) using the Pantheon sample \citep{2018ApJ...859..101S}, and BAO using results from the 6-degree Field Galaxy Survey and Sloan Digital Sky Survey Main Galaxy Sample \citep{2018MNRAS.481.2371C} as well as a compilation of different analyses of the BOSS DR12 data \citep{2017MNRAS.464.1168R, 2017MNRAS.466.2242B, 2018MNRAS.477.1153V}.
We refer the reader to Sections 5.1 and 5.2 of \citet{2020A&A...641A...6P} for more details on these additional probes.

In Table \ref{tab:cosmo_params}, we summarise the median and the 68\% confidence levels of the cosmological parameters constrained from our combined strong lensing models only (SL), CMB, BAO and SN, and their combinations.
To quantify the improvement on the figure of merit from the incorporation of the strong lensing constraint, we define the quantity $\delta i$ as the ratio between the 68\% confidence interval of the cosmological parameter $i$, obtained from the constraints without and with the strong lensing results.

For the flat-$\Lambda$CDM cosmology, the CMB-only constraints are very stringent compared to all other probes and dominate the combined posterior distributions.
Although the values of $\Omega_{\rm m}$ are consistent, the degeneracy between this parameter and the cluster mass distributions in the strong lensing models yields a relatively large statistical error.
In this simple cosmology, there are no improvements on the figure of merit when including the strong lensing information on the combined probes (see the values of $\bf \delta \Omega_{\rm m}$ on Table \ref{tab:cosmo_params}).
Nevertheless, considering models with additional free cosmological parameters, the strong lensing analyses add substantial constraints on the cosmological parameters.

The combined constraints for the flat-$w$CDM cosmology are shown in the right panel of Figure \ref{fig:flat_cosmo}, where the contours indicate the 68\% and 95\% confidence-level intervals.
Interestingly, strong lensing and CMB have an almost perpendicular parameter degeneracy, and the combination SL$\times$CMB improves significantly over the individual constraints.
The error bars on the parameters $\Omega_{\rm m}$ and $w$ are reduced by factors of 2.5 and 4.0, respectively (see Table \ref{tab:cosmo_params}).
Naturally, the improvement is minor in the case where the other three probes are combined, i.e. CMB$\times$BAO$\times$SN. However, the strong lensing information can still reduce the error bars by factors of $\approx 5\%$.

In the right panel of Figure \ref{fig:open_cosmo}, we show the combined constraints for the curved-$\Lambda$CDM cosmological model.
As discussed in the previous sections, cluster strong lensing is not very efficient in constraining the curvature of the Universe and provides a large error bar on this parameter.
However, it still improves the constraints on $\Omega_{\rm k}$ when combined with the CMB.
From Table \ref{tab:cosmo_params}, the figure of merit improves by a factor of $\approx 1.15$ on both $\Omega_{\rm m}$ and $\Omega_{\rm k}$ when both observable are considered.
When combining all observables, the improvements on $\Omega_{\rm m}$ is comparable to the flat-$w$CDM cosmology, but the constraint on $\Omega_{\rm k}$ is not affected.
Such effect is due to the large confidence region obtained from the strong lensing analyses (see both panels of Figure \ref{fig:open_cosmo}) that tends to overestimate the value of the curvature of the Universe.
The results from other probes, which indicate that the Universe is very close to being flat, and the weak constraining power on $\Omega_{\rm k}$ make cluster-strong-lensing cosmography not very efficient in probing this particular cosmological scenario.

For the CPL cosmology, the constraint from CMB only is not available in the public release from \citet{2020A&A...641A...6P} because of difficulties in the convergence of the posterior distribution.
Therefore, we first combine the strong-lensing constraints with the CMB$\times$BAO probes.
In addition to the 68\% confidence regions of the strong lensing analyses displayed in Figure \ref{fig:cpl}, we also show the 68\% and 95\% regions for the combined constraints.
The combined SL$\times$CMB$\times$BAO constraints improves the error bars on the three parameters $\Omega_{\rm m}$, $w_{\rm a}$ and $w_0$ by factors of $\approx 1.4-1.3$ (see Table \ref{tab:cosmo_params} for more details).
Although the parameter degeneracies of all probes have similar directions in the $w_{\rm a}$-$\Omega_{\rm m}$ plane, the probes have good complementarity in the other two projections.
In particular, high values of $w_0$ allowed by the CMB$\times$BAO probe are excluded by the strong lensing analyses.
We note that our results are also in excellent agreement when including constraints from SN (see the last row of Table {\ref{tab:cosmo_params}).
All these results demonstrate that cluster strong lensing cosmography can be used to constrain cosmological parameters and complement other standard probes.

\section{Conclusions}
\label{sec:conclusions}

In this work, we use galaxy cluster strong lenses to constrain the parameters of the background cosmology of the Universe, for the first time using a sample of galaxy clusters.
Thanks to the large number of multiple image families with spectroscopic redshifts, we are able to obtain competitive parameter constraints.
In order to estimate the efficiency of strong lensing cosmography in constraining cosmological parameters, we adopt four different cosmologies.
Moreover, we quantify the improvements when combining our posterior distributions with classical cosmological probes (i.e. CMB, BAO and SN).
Our main results are summarised as follows:

\begin{itemize}

\item We use the strong lensing models of each cluster to obtain the posterior distributions of the cosmological parameters. The combined constraints using the strong lensing models are all in agreement with other probes. Only for the curvature of the Universe ($\Omega_{\rm k}$) in the curved-$\Lambda$CDM model, the strong lensing constraints are not stringent. Thus, cluster strong lensing cosmography is not efficient in probing the curvature parameter $\Omega_{\rm k}$.

\item On the other hand, the strong lensing analyses provide stringent constraints on the dark energy equation of state parameters. In the case of a flat-$w$CDM cosmology, we obtain values of $\Omega_{\rm m} = 0.30_{-0.11}^{+0.09}$ and $w=-1.12_{-0.32}^{+0.17}$ for the 68\% confidence interval. The interval for the parameter $w$ is comparable to the CMB constraints (see Table \ref{tab:cosmo_params}).

\item For the flat-$w$CDM and CPL cosmologies, the strong lensing constraints on the equation of state for the dark energy component are consistent with the standard $\Lambda$CDM model (i.e. $w$ or $w_0$ = $-1$ and $w_{\rm a}=0$) within the 68\% confidence level. Even though the constraints on $w_0$ are weak, the parameter degeneracies in the CPL cosmology have some complementarity with other probes, especially in the projections $w_0$-$\Omega_{\rm m}$ and $w_0$-$w_{\rm a}$ (see Figure \ref{fig:cpl}).  

\item When combining strong lensing and CMB constraints, we find that we can improve the figure of merit on the cosmological parameters by significant factors. In the flat-$w$CDM cosmology, the constraints on $\Omega_{\rm m}$ and $w$ improve by factors of 2.5 and 4.0, respectively. For the curved-$\Lambda$CDM cosmology, the improvement is of the order of 1.15 in both $\Omega_{\rm m}$ and $\Omega_{\rm k}$ parameters. In the more complex CPL cosmological model, we combine the strong lensing posterior distributions with the CMB$\times$BAO constraints, leading to an improvement of $\approx 1.4 - 1.3$ on the three free parameters ($\Omega_{\rm m}$, $w_{0}$ and $w_{\rm a}$).

\item Finally, we find excellent agreement of our strong lensing analyses when comparing to the three ``classical'' probes (CMB, BAO and SN) altogether. For the parameter $\Omega_{\rm k}$ the combined constraint is not affected because strong lensing is weakly sensitive to this parameter. However, in all other cosmological models we show that cluster strong lensing cosmography can be a complementary probe and contribute to the combined probes.

\end{itemize}

Here we perform these analyses using a sample of five cluster strong lenses because of our knowledge of these systems from previous detailed works.
These cosmological constraints can be further improved by including new strong lensing models of additional clusters with similar data.
For instance, the number of clusters with deep spectroscopy, mainly using MUSE, is constantly increasing \citep[see e.g.][]{2021A&A...646A..83R} and these will be included in future analyses.

Moreover, several cosmological surveys from the ground and space, such as the Rubin Observatory Legacy Survey of Space and Time \citep[LSST,][]{2019ApJ...873..111I}, Euclid Space Telescope \citep{2011arXiv1110.3193L, 2021arXiv210801201S} and the Nancy Grace Roman Space Telescope \citep{2015arXiv150303757S}, will start to operate in the next few years. 
They will map most of the visible sky with unprecedented image quality and depth, and provide a large number of galaxy clusters that can be followed up with deep spectroscopy, increasing our sample of clusters by factors of tens and perhaps hundreds.
With data from these upcoming generation of telescopes, we might also be able to discern among different cosmological models.

Finally, we make publicly available the posterior distributions (in the format of parameter chains) of the cosmological and lens mass parameters obtained in this work.

\begin{acknowledgements}
GBC and SHS acknowledge the Max Planck Society for financial support through the Max Planck Research Group for SHS and the academic support from the German Centre for Cosmological Lensing.
CG and PR acknowledge financial support through grant PRIN-MIUR 2017WSCC32 ``Zooming into dark matter and proto-galaxies with massive lensing clusters'' (P.I.: P. Rosati).

This research made use of {\tt Astropy},\footnote{\url{http://www.astropy.org}} a community-developed core Python package for Astronomy \citep{astropy:2013, astropy:2018}, {\tt NumPy} \citep{harris2020array}, {\tt Matplotlib} \citep{Hunter:2007} and {\tt GetDist} \citep{2019arXiv191013970L}.
\end{acknowledgements}

\bibliographystyle{aa}
\bibliography{references}

\begin{thebibliography}{106}
\expandafter\ifx\csname natexlab\endcsname\relax\def\natexlab#1{#1}\fi

\bibitem[{{Abbott} {et~al.}(2018){Abbott}, {Abdalla}, {Alarcon}, {Aleksi{\'c}},
  {Allam}, {Allen}, {Amara}, {Annis}, {Asorey}, {Avila}, {Bacon}, {Balbinot},
  {Banerji}, {Banik}, {Barkhouse}, {Baumer}, {Baxter}, {Bechtol}, {Becker},
  {Benoit-L{\'e}vy}, {Benson}, {Bernstein}, {Bertin}, {Blazek}, {Bridle},
  {Brooks}, {Brout}, {Buckley-Geer}, {Burke}, {Busha}, {Campos}, {Capozzi},
  {Carnero Rosell}, {Carrasco Kind}, {Carretero}, {Castander}, {Cawthon},
  {Chang}, {Chen}, {Childress}, {Choi}, {Conselice}, {Crittenden}, {Crocce},
  {Cunha}, {D'Andrea}, {da Costa}, {Das}, {Davis}, {Davis}, {De Vicente},
  {DePoy}, {DeRose}, {Desai}, {Diehl}, {Dietrich}, {Dodelson}, {Doel},
  {Drlica-Wagner}, {Eifler}, {Elliott}, {Elsner}, {Elvin-Poole}, {Estrada},
  {Evrard}, {Fang}, {Fernandez}, {Fert{\'e}}, {Finley}, {Flaugher}, {Fosalba},
  {Friedrich}, {Frieman}, {Garc{\'\i}a-Bellido}, {Garcia-Fernandez}, {Gatti},
  {Gaztanaga}, {Gerdes}, {Giannantonio}, {Gill}, {Glazebrook}, {Goldstein},
  {Gruen}, {Gruendl}, {Gschwend}, {Gutierrez}, {Hamilton}, {Hartley}, {Hinton},
  {Honscheid}, {Hoyle}, {Huterer}, {Jain}, {James}, {Jarvis}, {Jeltema},
  {Johnson}, {Johnson}, {Kacprzak}, {Kent}, {Kim}, {King}, {Kirk}, {Kokron},
  {Kovacs}, {Krause}, {Krawiec}, {Kremin}, {Kuehn}, {Kuhlmann}, {Kuropatkin},
  {Lacasa}, {Lahav}, {Li}, {Liddle}, {Lidman}, {Lima}, {Lin}, {MacCrann},
  {Maia}, {Makler}, {Manera}, {March}, {Marshall}, {Martini}, {McMahon},
  {Melchior}, {Menanteau}, {Miquel}, {Miranda}, {Mudd}, {Muir}, {M{\"o}ller},
  {Neilsen}, {Nichol}, {Nord}, {Nugent}, {Ogando}, {Palmese}, {Peacock},
  {Peiris}, {Peoples}, {Percival}, {Petravick}, {Plazas}, {Porredon}, {Prat},
  {Pujol}, {Rau}, {Refregier}, {Ricker}, {Roe}, {Rollins}, {Romer}, {Roodman},
  {Rosenfeld}, {Ross}, {Rozo}, {Rykoff}, {Sako}, {Salvador}, {Samuroff},
  {S{\'a}nchez}, {Sanchez}, {Santiago}, {Scarpine}, {Schindler}, {Scolnic},
  {Secco}, {Serrano}, {Sevilla-Noarbe}, {Sheldon}, {Smith}, {Smith}, {Smith},
  {Soares-Santos}, {Sobreira}, {Suchyta}, {Tarle}, {Thomas}, {Troxel},
  {Tucker}, {Tucker}, {Uddin}, {Varga}, {Vielzeuf}, {Vikram}, {Vivas},
  {Walker}, {Wang}, {Wechsler}, {Weller}, {Wester}, {Wolf}, {Yanny}, {Yuan},
  {Zenteno}, {Zhang}, {Zhang}, {Zuntz}, \& {Dark Energy Survey
  Collaboration}}]{2018PhRvD..98d3526A}
{Abbott}, T.~M.~C., {Abdalla}, F.~B., {Alarcon}, A., {et~al.} 2018, \prd, 98,
  043526

\bibitem[{{Acebron} {et~al.}(2017){Acebron}, {Jullo}, {Limousin}, {Tilquin},
  {Giocoli}, {Jauzac}, {Mahler}, \& {Richard}}]{2017MNRAS.470.1809A}
{Acebron}, A., {Jullo}, E., {Limousin}, M., {et~al.} 2017, \mnras, 470, 1809

\bibitem[{{Astier} {et~al.}(2006){Astier}, {Guy}, {Regnault}, {Pain},
  {Aubourg}, {Balam}, {Basa}, {Carlberg}, {Fabbro}, {Fouchez}, {Hook},
  {Howell}, {Lafoux}, {Neill}, {Palanque-Delabrouille}, {Perrett}, {Pritchet},
  {Rich}, {Sullivan}, {Taillet}, {Aldering}, {Antilogus}, {Arsenijevic},
  {Balland}, {Baumont}, {Bronder}, {Courtois}, {Ellis}, {Filiol},
  {Gon{\c{c}}alves}, {Goobar}, {Guide}, {Hardin}, {Lusset}, {Lidman},
  {McMahon}, {Mouchet}, {Mourao}, {Perlmutter}, {Ripoche}, {Tao}, \&
  {Walton}}]{2006A&A...447...31A}
{Astier}, P., {Guy}, J., {Regnault}, N., {et~al.} 2006, \aap, 447, 31

\bibitem[{{Astropy Collaboration} {et~al.}(2018){Astropy Collaboration},
  {Price-Whelan}, {Sip{\H{o}}cz}, {G{\"u}nther}, {Lim}, {Crawford}, {Conseil},
  {Shupe}, {Craig}, {Dencheva}, {Ginsburg}, {Vand erPlas}, {Bradley},
  {P{\'e}rez-Su{\'a}rez}, {de Val-Borro}, {Aldcroft}, {Cruz}, {Robitaille},
  {Tollerud}, {Ardelean}, {Babej}, {Bach}, {Bachetti}, {Bakanov}, {Bamford},
  {Barentsen}, {Barmby}, {Baumbach}, {Berry}, {Biscani}, {Boquien}, {Bostroem},
  {Bouma}, {Brammer}, {Bray}, {Breytenbach}, {Buddelmeijer}, {Burke},
  {Calderone}, {Cano Rodr{\'\i}guez}, {Cara}, {Cardoso}, {Cheedella}, {Copin},
  {Corrales}, {Crichton}, {D'Avella}, {Deil}, {Depagne}, {Dietrich}, {Donath},
  {Droettboom}, {Earl}, {Erben}, {Fabbro}, {Ferreira}, {Finethy}, {Fox},
  {Garrison}, {Gibbons}, {Goldstein}, {Gommers}, {Greco}, {Greenfield},
  {Groener}, {Grollier}, {Hagen}, {Hirst}, {Homeier}, {Horton}, {Hosseinzadeh},
  {Hu}, {Hunkeler}, {Ivezi{\'c}}, {Jain}, {Jenness}, {Kanarek}, {Kendrew},
  {Kern}, {Kerzendorf}, {Khvalko}, {King}, {Kirkby}, {Kulkarni}, {Kumar},
  {Lee}, {Lenz}, {Littlefair}, {Ma}, {Macleod}, {Mastropietro}, {McCully},
  {Montagnac}, {Morris}, {Mueller}, {Mumford}, {Muna}, {Murphy}, {Nelson},
  {Nguyen}, {Ninan}, {N{\"o}the}, {Ogaz}, {Oh}, {Parejko}, {Parley}, {Pascual},
  {Patil}, {Patil}, {Plunkett}, {Prochaska}, {Rastogi}, {Reddy Janga},
  {Sabater}, {Sakurikar}, {Seifert}, {Sherbert}, {Sherwood-Taylor}, {Shih},
  {Sick}, {Silbiger}, {Singanamalla}, {Singer}, {Sladen}, {Sooley},
  {Sornarajah}, {Streicher}, {Teuben}, {Thomas}, {Tremblay}, {Turner},
  {Terr{\'o}n}, {van Kerkwijk}, {de la Vega}, {Watkins}, {Weaver}, {Whitmore},
  {Woillez}, {Zabalza}, \& {Astropy Contributors}}]{astropy:2018}
{Astropy Collaboration}, {Price-Whelan}, A.~M., {Sip{\H{o}}cz}, B.~M., {et~al.}
  2018, \aj, 156, 123

\bibitem[{{Astropy Collaboration} {et~al.}(2013){Astropy Collaboration},
  {Robitaille}, {Tollerud}, {Greenfield}, {Droettboom}, {Bray}, {Aldcroft},
  {Davis}, {Ginsburg}, {Price-Whelan}, {Kerzendorf}, {Conley}, {Crighton},
  {Barbary}, {Muna}, {Ferguson}, {Grollier}, {Parikh}, {Nair}, {Unther},
  {Deil}, {Woillez}, {Conseil}, {Kramer}, {Turner}, {Singer}, {Fox}, {Weaver},
  {Zabalza}, {Edwards}, {Azalee Bostroem}, {Burke}, {Casey}, {Crawford},
  {Dencheva}, {Ely}, {Jenness}, {Labrie}, {Lim}, {Pierfederici}, {Pontzen},
  {Ptak}, {Refsdal}, {Servillat}, \& {Streicher}}]{astropy:2013}
{Astropy Collaboration}, {Robitaille}, T.~P., {Tollerud}, E.~J., {et~al.} 2013,
  \aap, 558, A33

\bibitem[{{Bacon} {et~al.}(2014){Bacon}, {Vernet}, {Borisova}, {Bouch{\'e}},
  {Brinchmann}, {Carollo}, {Carton}, {Caruana}, {Cerda}, {Contini}, {Franx},
  {Girard}, {Guerou}, {Haddad}, {Hau}, {Herenz}, {Herrera}, {Husemann},
  {Husser}, {Jarno}, {Kamann}, {Krajnovic}, {Lilly}, {Mainieri}, {Martinsson},
  {Palsa}, {Patricio}, {P{\'e}contal}, {Pello}, {Piqueras}, {Richard},
  {Sandin}, {Schroetter}, {Selman}, {Shirazi}, {Smette}, {Soto}, {Streicher},
  {Urrutia}, {Weilbacher}, {Wisotzki}, \& {Zins}}]{2014Msngr.157...13B}
{Bacon}, R., {Vernet}, J., {Borisova}, E., {et~al.} 2014, The Messenger, 157,
  13

\bibitem[{Beauchesne {et~al.}(2021)Beauchesne, Clément, Richard, Kneib, \&
  }]{beauchesne2021improving}
Beauchesne, B., Clément, B., Richard, J., Kneib, J.-P., \& . 2021, Improving
  parametric mass modelling of lensing clusters through a perturbative approach

\bibitem[{{Belli} {et~al.}(2013){Belli}, {Jones}, {Ellis}, \&
  {Richard}}]{2013ApJ...772..141B}
{Belli}, S., {Jones}, T., {Ellis}, R.~S., \& {Richard}, J. 2013, \apj, 772, 141

\bibitem[{{Bergamini} {et~al.}(2021){Bergamini}, {Rosati}, {Vanzella},
  {Caminha}, {Grillo}, {Mercurio}, {Meneghetti}, {Angora}, {Calura}, {Nonino},
  \& {Tozzi}}]{2021A&A...645A.140B}
{Bergamini}, P., {Rosati}, P., {Vanzella}, E., {et~al.} 2021, \aap, 645, A140

\bibitem[{{Beutler} {et~al.}(2017){Beutler}, {Seo}, {Saito}, {Chuang},
  {Cuesta}, {Eisenstein}, {Gil-Mar{\'\i}n}, {Grieb}, {Hand}, {Kitaura}, {Modi},
  {Nichol}, {Olmstead}, {Percival}, {Prada}, {S{\'a}nchez}, {Rodriguez-Torres},
  {Ross}, {Ross}, {Schneider}, {Tinker}, {Tojeiro}, \&
  {Vargas-Maga{\~n}a}}]{2017MNRAS.466.2242B}
{Beutler}, F., {Seo}, H.-J., {Saito}, S., {et~al.} 2017, \mnras, 466, 2242

\bibitem[{{Birrer} {et~al.}(2020){Birrer}, {Shajib}, {Galan}, {Millon}, {Treu},
  {Agnello}, {Auger}, {Chen}, {Christensen}, {Collett}, {Courbin}, {Fassnacht},
  {Koopmans}, {Marshall}, {Park}, {Rusu}, {Sluse}, {Spiniello}, {Suyu},
  {Wagner-Carena}, {Wong}, {Barnab{\`e}}, {Bolton}, {Czoske}, {Ding},
  {Frieman}, \& {Van de Vyvere}}]{2020A&A...643A.165B}
{Birrer}, S., {Shajib}, A.~J., {Galan}, A., {et~al.} 2020, \aap, 643, A165

\bibitem[{{Blandford} \& {Narayan}(1992)}]{1992ARA&A..30..311B}
{Blandford}, R.~D. \& {Narayan}, R. 1992, \araa, 30, 311

\bibitem[{Brooks \& Gelman(1998)}]{Brooks}
Brooks, S.~P. \& Gelman, A. 1998, Journal of Computational and Graphical
  Statistics, 7, 434

\bibitem[{{Caminha} {et~al.}(2016{\natexlab{a}}){Caminha}, {Grillo}, {Rosati},
  {Balestra}, {Karman}, {Lombardi}, {Mercurio}, {Nonino}, {Tozzi}, {Zitrin},
  {Biviano}, {Girardi}, {Koekemoer}, {Melchior}, {Meneghetti}, {Munari},
  {Suyu}, {Umetsu}, {Annunziatella}, {Borgani}, {Broadhurst}, {Caputi}, {Coe},
  {Delgado-Correal}, {Ettori}, {Fritz}, {Frye}, {Gobat}, {Maier}, {Monna},
  {Postman}, {Sartoris}, {Seitz}, {Vanzella}, \&
  {Ziegler}}]{2016A&A...587A..80C}
{Caminha}, G.~B., {Grillo}, C., {Rosati}, P., {et~al.} 2016{\natexlab{a}},
  \aap, 587, A80

\bibitem[{{Caminha} {et~al.}(2017{\natexlab{a}}){Caminha}, {Grillo}, {Rosati},
  {Balestra}, {Mercurio}, {Vanzella}, {Biviano}, {Caputi}, {Delgado-Correal},
  {Karman}, {Lombardi}, {Meneghetti}, {Sartoris}, \&
  {Tozzi}}]{2017A&A...600A..90C}
{Caminha}, G.~B., {Grillo}, C., {Rosati}, P., {et~al.} 2017{\natexlab{a}},
  \aap, 600, A90

\bibitem[{{Caminha} {et~al.}(2017{\natexlab{b}}){Caminha}, {Grillo}, {Rosati},
  {Meneghetti}, {Mercurio}, {Ettori}, {Balestra}, {Biviano}, {Umetsu},
  {Vanzella}, {Annunziatella}, {Bonamigo}, {Delgado-Correal}, {Girardi},
  {Lombardi}, {Nonino}, {Sartoris}, {Tozzi}, {Bartelmann}, {Bradley}, {Caputi},
  {Coe}, {Ford}, {Fritz}, {Gobat}, {Postman}, {Seitz}, \&
  {Zitrin}}]{2017A&A...607A..93C}
{Caminha}, G.~B., {Grillo}, C., {Rosati}, P., {et~al.} 2017{\natexlab{b}},
  \aap, 607, A93

\bibitem[{{Caminha} {et~al.}(2016{\natexlab{b}}){Caminha}, {Karman}, {Rosati},
  {Caputi}, {Arrigoni Battaia}, {Balestra}, {Grillo}, {Mercurio}, {Nonino}, \&
  {Vanzella}}]{2016A&A...595A.100C}
{Caminha}, G.~B., {Karman}, W., {Rosati}, P., {et~al.} 2016{\natexlab{b}},
  \aap, 595, A100

\bibitem[{{Caminha} {et~al.}(2019){Caminha}, {Rosati}, {Grillo}, {Rosani},
  {Caputi}, {Meneghetti}, {Mercurio}, {Balestra}, {Bergamini}, {Biviano},
  {Nonino}, {Umetsu}, {Vanzella}, {Annunziatella}, {Broadhurst},
  {Delgado-Correal}, {Demarco}, {Koekemoer}, {Lombardi}, {Maier}, {Verdugo}, \&
  {Zitrin}}]{2019A&A...632A..36C}
{Caminha}, G.~B., {Rosati}, P., {Grillo}, C., {et~al.} 2019, \aap, 632, A36

\bibitem[{{Cao} {et~al.}(2012){Cao}, {Pan}, {Biesiada}, {Godlowski}, \&
  {Zhu}}]{2012JCAP...03..016C}
{Cao}, S., {Pan}, Y., {Biesiada}, M., {Godlowski}, W., \& {Zhu}, Z.-H. 2012,
  \jcap, 2012, 016

\bibitem[{{Carlsten} {et~al.}(2020){Carlsten}, {Greene}, {Peter}, {Greco}, \&
  {Beaton}}]{2020ApJ...902..124C}
{Carlsten}, S.~G., {Greene}, J.~E., {Peter}, A. H.~G., {Greco}, J.~P., \&
  {Beaton}, R.~L. 2020, \apj, 902, 124

\bibitem[{{Carter} {et~al.}(2018){Carter}, {Beutler}, {Percival}, {Blake},
  {Koda}, \& {Ross}}]{2018MNRAS.481.2371C}
{Carter}, P., {Beutler}, F., {Percival}, W.~J., {et~al.} 2018, \mnras, 481,
  2371

\bibitem[{{Chen} {et~al.}(2019){Chen}, {Fassnacht}, {Suyu}, {Rusu}, {Chan},
  {Wong}, {Auger}, {Hilbert}, {Bonvin}, {Birrer}, {Millon}, {Koopmans},
  {Lagattuta}, {McKean}, {Vegetti}, {Courbin}, {Ding}, {Halkola}, {Jee},
  {Shajib}, {Sluse}, {Sonnenfeld}, \& {Treu}}]{2019MNRAS.490.1743C}
{Chen}, G. C.~F., {Fassnacht}, C.~D., {Suyu}, S.~H., {et~al.} 2019, \mnras,
  490, 1743

\bibitem[{{Chevallier} \& {Polarski}(2001)}]{2001IJMPD..10..213C}
{Chevallier}, M. \& {Polarski}, D. 2001, International Journal of Modern
  Physics D, 10, 213

\bibitem[{{Collett} \& {Auger}(2014)}]{2014MNRAS.443..969C}
{Collett}, T.~E. \& {Auger}, M.~W. 2014, \mnras, 443, 969

\bibitem[{{Courbin} {et~al.}(2018){Courbin}, {Bonvin}, {Buckley-Geer},
  {Fassnacht}, {Frieman}, {Lin}, {Marshall}, {Suyu}, {Treu}, {Anguita},
  {Motta}, {Meylan}, {Paic}, {Tewes}, {Agnello}, {Chao}, {Chijani}, {Gilman},
  {Rojas}, {Williams}, {Hempel}, {Kim}, {Lachaume}, {Rabus}, {Abbott}, {Allam},
  {Annis}, {Banerji}, {Bechtol}, {Benoit-L{\'e}vy}, {Brooks}, {Burke}, {Carnero
  Rosell}, {Carrasco Kind}, {Carretero}, {D'Andrea}, {da Costa}, {Davis},
  {DePoy}, {Desai}, {Flaugher}, {Fosalba}, {Garc{\'\i}a-Bellido}, {Gaztanaga},
  {Goldstein}, {Gruen}, {Gruendl}, {Gschwend}, {Gutierrez}, {Honscheid},
  {James}, {Kuehn}, {Kuhlmann}, {Kuropatkin}, {Lahav}, {Lima}, {Maia}, {March},
  {Marshall}, {McMahon}, {Menanteau}, {Miquel}, {Nord}, {Plazas}, {Sanchez},
  {Scarpine}, {Schindler}, {Schubnell}, {Sevilla-Noarbe}, {Smith},
  {Soares-Santos}, {Sobreira}, {Suchyta}, {Tarle}, {Tucker}, {Walker}, \&
  {Wester}}]{2018A&A...609A..71C}
{Courbin}, F., {Bonvin}, V., {Buckley-Geer}, E., {et~al.} 2018, \aap, 609, A71

\bibitem[{{D'Aloisio} \& {Natarajan}(2011)}]{2011MNRAS.411.1628D}
{D'Aloisio}, A. \& {Natarajan}, P. 2011, \mnras, 411, 1628

\bibitem[{{Dawson} {et~al.}(2013){Dawson}, {Schlegel}, {Ahn}, {Anderson},
  {Aubourg}, {Bailey}, {Barkhouser}, {Bautista}, {Beifiori}, {Berlind},
  {Bhardwaj}, {Bizyaev}, {Blake}, {Blanton}, {Blomqvist}, {Bolton}, {Borde},
  {Bovy}, {Brandt}, {Brewington}, {Brinkmann}, {Brown}, {Brownstein}, {Bundy},
  {Busca}, {Carithers}, {Carnero}, {Carr}, {Chen}, {Comparat}, {Connolly},
  {Cope}, {Croft}, {Cuesta}, {da Costa}, {Davenport}, {Delubac}, {de Putter},
  {Dhital}, {Ealet}, {Ebelke}, {Eisenstein}, {Escoffier}, {Fan}, {Filiz Ak},
  {Finley}, {Font-Ribera}, {G{\'e}nova-Santos}, {Gunn}, {Guo}, {Haggard},
  {Hall}, {Hamilton}, {Harris}, {Harris}, {Ho}, {Hogg}, {Holder}, {Honscheid},
  {Huehnerhoff}, {Jordan}, {Jordan}, {Kauffmann}, {Kazin}, {Kirkby}, {Klaene},
  {Kneib}, {Le Goff}, {Lee}, {Long}, {Loomis}, {Lundgren}, {Lupton}, {Maia},
  {Makler}, {Malanushenko}, {Malanushenko}, {Mandelbaum}, {Manera}, {Maraston},
  {Margala}, {Masters}, {McBride}, {McDonald}, {McGreer}, {McMahon}, {Mena},
  {Miralda-Escud{\'e}}, {Montero-Dorta}, {Montesano}, {Muna}, {Myers},
  {Naugle}, {Nichol}, {Noterdaeme}, {Nuza}, {Olmstead}, {Oravetz}, {Oravetz},
  {Owen}, {Padmanabhan}, {Palanque-Delabrouille}, {Pan}, {Parejko},
  {P{\^a}ris}, {Percival}, {P{\'e}rez-Fournon}, {P{\'e}rez-R{\`a}fols},
  {Petitjean}, {Pfaffenberger}, {Pforr}, {Pieri}, {Prada}, {Price-Whelan},
  {Raddick}, {Rebolo}, {Rich}, {Richards}, {Rockosi}, {Roe}, {Ross}, {Ross},
  {Rossi}, {Rubi{\~n}o-Martin}, {Samushia}, {S{\'a}nchez}, {Sayres}, {Schmidt},
  {Schneider}, {Sc{\'o}ccola}, {Seo}, {Shelden}, {Sheldon}, {Shen}, {Shu},
  {Slosar}, {Smee}, {Snedden}, {Stauffer}, {Steele}, {Strauss}, {Streblyanska},
  {Suzuki}, {Swanson}, {Tal}, {Tanaka}, {Thomas}, {Tinker}, {Tojeiro},
  {Tremonti}, {Vargas Maga{\~n}a}, {Verde}, {Viel}, {Wake}, {Watson}, {Weaver},
  {Weinberg}, {Weiner}, {West}, {White}, {Wood-Vasey}, {Yeche}, {Zehavi},
  {Zhao}, \& {Zheng}}]{2013AJ....145...10D}
{Dawson}, K.~S., {Schlegel}, D.~J., {Ahn}, C.~P., {et~al.} 2013, \aj, 145, 10

\bibitem[{{DES Collaboration} {et~al.}(2021){DES Collaboration}, {Abbott},
  {Aguena}, {Alarcon}, {Allam}, {Alves}, {Amon}, {Andrade-Oliveira}, {Annis},
  {Avila}, {Bacon}, {Baxter}, {Bechtol}, {Becker}, {Bernstein}, {Bhargava},
  {Birrer}, {Blazek}, {Brandao-Souza}, {Bridle}, {Brooks}, {Buckley-Geer},
  {Burke}, {Camacho}, {Campos}, {Carnero Rosell}, {Carrasco Kind}, {Carretero},
  {Castander}, {Cawthon}, {Chang}, {Chen}, {Chen}, {Choi}, {Conselice},
  {Cordero}, {Costanzi}, {Crocce}, {da Costa}, {da Silva Pereira}, {Davis},
  {Davis}, {De Vicente}, {DeRose}, {Desai}, {Di Valentino}, {Diehl},
  {Dietrich}, {Dodelson}, {Doel}, {Doux}, {Drlica-Wagner}, {Eckert}, {Eifler},
  {Elsner}, {Elvin-Poole}, {Everett}, {Evrard}, {Fang}, {Farahi}, {Fernandez},
  {Ferrero}, {Fert{\'e}}, {Fosalba}, {Friedrich}, {Frieman},
  {Garc{\'\i}a-Bellido}, {Gatti}, {Gaztanaga}, {Gerdes}, {Giannantonio},
  {Giannini}, {Gruen}, {Gruendl}, {Gschwend}, {Gutierrez}, {Harrison},
  {Hartley}, {Herner}, {Hinton}, {Hollowood}, {Honscheid}, {Hoyle}, {Huff},
  {Huterer}, {Jain}, {James}, {Jarvis}, {Jeffrey}, {Jeltema}, {Kovacs},
  {Krause}, {Kron}, {Kuehn}, {Kuropatkin}, {Lahav}, {Leget}, {Lemos}, {Liddle},
  {Lidman}, {Lima}, {Lin}, {MacCrann}, {Maia}, {Marshall}, {Martini},
  {McCullough}, {Melchior}, {Mena-Fern{\'a}ndez}, {Menanteau}, {Miquel},
  {Mohr}, {Morgan}, {Muir}, {Myles}, {Nadathur}, {Navarro-Alsina}, {Nichol},
  {Ogando}, {Omori}, {Palmese}, {Pandey}, {Park}, {Paz-Chinch{\'o}n},
  {Petravick}, {Pieres}, {Plazas Malag{\'o}n}, {Porredon}, {Prat}, {Raveri},
  {Rodriguez-Monroy}, {Rollins}, {Romer}, {Roodman}, {Rosenfeld}, {Ross},
  {Rykoff}, {Samuroff}, {S{\'a}nchez}, {Sanchez}, {Sanchez}, {Sanchez Cid},
  {Scarpine}, {Schubnell}, {Scolnic}, {Secco}, {Serrano}, {Sevilla-Noarbe},
  {Sheldon}, {Shin}, {Smith}, {Soares-Santos}, {Suchyta}, {Swanson}, {Tabbutt},
  {Tarle}, {Thomas}, {To}, {Troja}, {Troxel}, {Tucker}, {Tutusaus}, {Varga},
  {Walker}, {Weaverdyck}, {Weller}, {Yanny}, {Yin}, {Zhang}, \&
  {Zuntz}}]{2021arXiv210513549D}
{DES Collaboration}, {Abbott}, T.~M.~C., {Aguena}, M., {et~al.} 2021, arXiv
  e-prints, arXiv:2105.13549

\bibitem[{{Efstathiou} {et~al.}(2002){Efstathiou}, {Moody}, {Peacock},
  {Percival}, {Baugh}, {Bland-Hawthorn}, {Bridges}, {Cannon}, {Cole},
  {Colless}, {Collins}, {Couch}, {Dalton}, {de Propris}, {Driver}, {Ellis},
  {Frenk}, {Glazebrook}, {Jackson}, {Lahav}, {Lewis}, {Lumsden}, {Maddox},
  {Norberg}, {Peterson}, {Sutherland}, \& {Taylor}}]{2002MNRAS.330L..29E}
{Efstathiou}, G., {Moody}, S., {Peacock}, J.~A., {et~al.} 2002, \mnras, 330,
  L29

\bibitem[{{Eisenstein} {et~al.}(2005){Eisenstein}, {Zehavi}, {Hogg},
  {Scoccimarro}, {Blanton}, {Nichol}, {Scranton}, {Seo}, {Tegmark}, {Zheng},
  {Anderson}, {Annis}, {Bahcall}, {Brinkmann}, {Burles}, {Castander},
  {Connolly}, {Csabai}, {Doi}, {Fukugita}, {Frieman}, {Glazebrook}, {Gunn},
  {Hendry}, {Hennessy}, {Ivezi{\'c}}, {Kent}, {Knapp}, {Lin}, {Loh}, {Lupton},
  {Margon}, {McKay}, {Meiksin}, {Munn}, {Pope}, {Richmond}, {Schlegel},
  {Schneider}, {Shimasaku}, {Stoughton}, {Strauss}, {SubbaRao}, {Szalay},
  {Szapudi}, {Tucker}, {Yanny}, \& {York}}]{2005ApJ...633..560E}
{Eisenstein}, D.~J., {Zehavi}, I., {Hogg}, D.~W., {et~al.} 2005, \apj, 633, 560

\bibitem[{{El{\'{\i}}asd{\'o}ttir} {et~al.}(2007){El{\'{\i}}asd{\'o}ttir},
  {Limousin}, {Richard}, {Hjorth}, {Kneib}, {Natarajan}, {Pedersen}, {Jullo},
  \& {Paraficz}}]{2007arXiv0710.5636E}
{El{\'{\i}}asd{\'o}ttir}, {\'A}., {Limousin}, M., {Richard}, J., {et~al.} 2007,
  ArXiv e-prints [\eprint[arXiv]{0710.5636}]

\bibitem[{{Fassnacht} {et~al.}(2002){Fassnacht}, {Xanthopoulos}, {Koopmans}, \&
  {Rusin}}]{2002ApJ...581..823F}
{Fassnacht}, C.~D., {Xanthopoulos}, E., {Koopmans}, L.~V.~E., \& {Rusin}, D.
  2002, \apj, 581, 823

\bibitem[{{Gilmore} \& {Natarajan}(2009)}]{2009MNRAS.396..354G}
{Gilmore}, J. \& {Natarajan}, P. 2009, \mnras, 396, 354

\bibitem[{{Gnedin} {et~al.}(2011){Gnedin}, {Ceverino}, {Gnedin}, {Klypin},
  {Kravtsov}, {Levine}, {Nagai}, \& {Yepes}}]{2011arXiv1108.5736G}
{Gnedin}, O.~Y., {Ceverino}, D., {Gnedin}, N.~Y., {et~al.} 2011, ArXiv e-prints
  [\eprint[arXiv]{1108.5736}]

\bibitem[{{Gnedin} {et~al.}(2004){Gnedin}, {Kravtsov}, {Klypin}, \&
  {Nagai}}]{2004ApJ...616...16G}
{Gnedin}, O.~Y., {Kravtsov}, A.~V., {Klypin}, A.~A., \& {Nagai}, D. 2004, \apj,
  616, 16

\bibitem[{{Golse} {et~al.}(2002){Golse}, {Kneib}, \&
  {Soucail}}]{2002A&A...387..788G}
{Golse}, G., {Kneib}, J.~P., \& {Soucail}, G. 2002, \aap, 387, 788

\bibitem[{{Grillo} {et~al.}(2016){Grillo}, {Karman}, {Suyu}, {Rosati},
  {Balestra}, {Mercurio}, {Lombardi}, {Treu}, {Caminha}, {Halkola}, {Rodney},
  {Gavazzi}, \& {Caputi}}]{2016ApJ...822...78G}
{Grillo}, C., {Karman}, W., {Suyu}, S.~H., {et~al.} 2016, \apj, 822, 78

\bibitem[{{Grillo} {et~al.}(2008){Grillo}, {Lombardi}, \&
  {Bertin}}]{2008A&A...477..397G}
{Grillo}, C., {Lombardi}, M., \& {Bertin}, G. 2008, \aap, 477, 397

\bibitem[{{Grillo} {et~al.}(2018){Grillo}, {Rosati}, {Suyu}, {Balestra},
  {Caminha}, {Halkola}, {Kelly}, {Lombardi}, {Mercurio}, {Rodney}, \&
  {Treu}}]{2018ApJ...860...94G}
{Grillo}, C., {Rosati}, P., {Suyu}, S.~H., {et~al.} 2018, \apj, 860, 94

\bibitem[{{Grillo} {et~al.}(2020){Grillo}, {Rosati}, {Suyu}, {Caminha},
  {Mercurio}, \& {Halkola}}]{2020ApJ...898...87G}
{Grillo}, C., {Rosati}, P., {Suyu}, S.~H., {et~al.} 2020, \apj, 898, 87

\bibitem[{{Grillo} {et~al.}(2015){Grillo}, {Suyu}, {Rosati}, {Mercurio},
  {Balestra}, {Munari}, {Nonino}, {Caminha}, {Lombardi}, {De Lucia}, {Borgani},
  {Gobat}, {Biviano}, {Girardi}, {Umetsu}, {Coe}, {Koekemoer}, {Postman},
  {Zitrin}, {Halkola}, {Broadhurst}, {Sartoris}, {Presotto}, {Annunziatella},
  {Maier}, {Fritz}, {Vanzella}, \& {Frye}}]{2015ApJ...800...38G}
{Grillo}, C., {Suyu}, S.~H., {Rosati}, P., {et~al.} 2015, \apj, 800, 38

\bibitem[{Harris {et~al.}(2020)Harris, Millman, van~der Walt, Gommers,
  Virtanen, Cournapeau, Wieser, Taylor, Berg, Smith, Kern, Picus, Hoyer, van
  Kerkwijk, Brett, Haldane, del R{\'{i}}o, Wiebe, Peterson,
  G{\'{e}}rard-Marchant, Sheppard, Reddy, Weckesser, Abbasi, Gohlke, \&
  Oliphant}]{harris2020array}
Harris, C.~R., Millman, K.~J., van~der Walt, S.~J., {et~al.} 2020, Nature, 585,
  357

\bibitem[{{Hinshaw} {et~al.}(2013){Hinshaw}, {Larson}, {Komatsu}, {Spergel},
  {Bennett}, {Dunkley}, {Nolta}, {Halpern}, {Hill}, {Odegard}, {Page}, {Smith},
  {Weiland}, {Gold}, {Jarosik}, {Kogut}, {Limon}, {Meyer}, {Tucker}, {Wollack},
  \& {Wright}}]{2013ApJS..208...19H}
{Hinshaw}, G., {Larson}, D., {Komatsu}, E., {et~al.} 2013, \apjs, 208, 19

\bibitem[{{Host}(2012)}]{2012MNRAS.420L..18H}
{Host}, O. 2012, \mnras, 420, L18

\bibitem[{{Huang} {et~al.}(2016){Huang}, {Lemaux}, {Schmidt}, {Hoag},
  {Brada{\v{c}}}, {Treu}, {Dijkstra}, {Fontana}, {Henry}, {Malkan}, {Mason},
  {Morishita}, {Pentericci}, {Ryan}, {Trenti}, \& {Wang}}]{2016ApJ...823L..14H}
{Huang}, K.-H., {Lemaux}, B.~C., {Schmidt}, K.~B., {et~al.} 2016, \apjl, 823,
  L14

\bibitem[{Hunter(2007)}]{Hunter:2007}
Hunter, J.~D. 2007, Computing in Science \& Engineering, 9, 90

\bibitem[{{Ivezi{\'c}} {et~al.}(2019){Ivezi{\'c}}, {Kahn}, {Tyson}, {Abel},
  {Acosta}, {Allsman}, {Alonso}, {AlSayyad}, {Anderson}, {Andrew}, {Angel},
  {Angeli}, {Ansari}, {Antilogus}, {Araujo}, {Armstrong}, {Arndt}, {Astier},
  {Aubourg}, {Auza}, {Axelrod}, {Bard}, {Barr}, {Barrau}, {Bartlett}, {Bauer},
  {Bauman}, {Baumont}, {Bechtol}, {Bechtol}, {Becker}, {Becla}, {Beldica},
  {Bellavia}, {Bianco}, {Biswas}, {Blanc}, {Blazek}, {Blandford}, {Bloom},
  {Bogart}, {Bond}, {Booth}, {Borgland}, {Borne}, {Bosch}, {Boutigny},
  {Brackett}, {Bradshaw}, {Brandt}, {Brown}, {Bullock}, {Burchat}, {Burke},
  {Cagnoli}, {Calabrese}, {Callahan}, {Callen}, {Carlin}, {Carlson},
  {Chandrasekharan}, {Charles-Emerson}, {Chesley}, {Cheu}, {Chiang}, {Chiang},
  {Chirino}, {Chow}, {Ciardi}, {Claver}, {Cohen-Tanugi}, {Cockrum}, {Coles},
  {Connolly}, {Cook}, {Cooray}, {Covey}, {Cribbs}, {Cui}, {Cutri}, {Daly},
  {Daniel}, {Daruich}, {Daubard}, {Daues}, {Dawson}, {Delgado}, {Dellapenna},
  {de Peyster}, {de Val-Borro}, {Digel}, {Doherty}, {Dubois},
  {Dubois-Felsmann}, {Durech}, {Economou}, {Eifler}, {Eracleous}, {Emmons},
  {Fausti Neto}, {Ferguson}, {Figueroa}, {Fisher-Levine}, {Focke}, {Foss},
  {Frank}, {Freemon}, {Gangler}, {Gawiser}, {Geary}, {Gee}, {Geha}, {Gessner},
  {Gibson}, {Gilmore}, {Glanzman}, {Glick}, {Goldina}, {Goldstein}, {Goodenow},
  {Graham}, {Gressler}, {Gris}, {Guy}, {Guyonnet}, {Haller}, {Harris},
  {Hascall}, {Haupt}, {Hernandez}, {Herrmann}, {Hileman}, {Hoblitt}, {Hodgson},
  {Hogan}, {Howard}, {Huang}, {Huffer}, {Ingraham}, {Innes}, {Jacoby}, {Jain},
  {Jammes}, {Jee}, {Jenness}, {Jernigan}, {Jevremovi{\'c}}, {Johns}, {Johnson},
  {Johnson}, {Jones}, {Juramy-Gilles}, {Juri{\'c}}, {Kalirai}, {Kallivayalil},
  {Kalmbach}, {Kantor}, {Karst}, {Kasliwal}, {Kelly}, {Kessler}, {Kinnison},
  {Kirkby}, {Knox}, {Kotov}, {Krabbendam}, {Krughoff}, {Kub{\'a}nek},
  {Kuczewski}, {Kulkarni}, {Ku}, {Kurita}, {Lage}, {Lambert}, {Lange},
  {Langton}, {Le Guillou}, {Levine}, {Liang}, {Lim}, {Lintott}, {Long},
  {Lopez}, {Lotz}, {Lupton}, {Lust}, {MacArthur}, {Mahabal}, {Mandelbaum},
  {Markiewicz}, {Marsh}, {Marshall}, {Marshall}, {May}, {McKercher}, {McQueen},
  {Meyers}, {Migliore}, {Miller}, {Mills}, {Miraval}, {Moeyens}, {Moolekamp},
  {Monet}, {Moniez}, {Monkewitz}, {Montgomery}, {Morrison}, {Mueller},
  {Muller}, {Mu{\~n}oz Arancibia}, {Neill}, {Newbry}, {Nief}, {Nomerotski},
  {Nordby}, {O'Connor}, {Oliver}, {Olivier}, {Olsen}, {O'Mullane}, {Ortiz},
  {Osier}, {Owen}, {Pain}, {Palecek}, {Parejko}, {Parsons}, {Pease},
  {Peterson}, {Peterson}, {Petravick}, {Libby Petrick}, {Petry},
  {Pierfederici}, {Pietrowicz}, {Pike}, {Pinto}, {Plante}, {Plate}, {Plutchak},
  {Price}, {Prouza}, {Radeka}, {Rajagopal}, {Rasmussen}, {Regnault}, {Reil},
  {Reiss}, {Reuter}, {Ridgway}, {Riot}, {Ritz}, {Robinson}, {Roby}, {Roodman},
  {Rosing}, {Roucelle}, {Rumore}, {Russo}, {Saha}, {Sassolas}, {Schalk},
  {Schellart}, {Schindler}, {Schmidt}, {Schneider}, {Schneider}, {Schoening},
  {Schumacher}, {Schwamb}, {Sebag}, {Selvy}, {Sembroski}, {Seppala}, {Serio},
  {Serrano}, {Shaw}, {Shipsey}, {Sick}, {Silvestri}, {Slater}, {Smith},
  {Smith}, {Sobhani}, {Soldahl}, {Storrie-Lombardi}, {Stover}, {Strauss},
  {Street}, {Stubbs}, {Sullivan}, {Sweeney}, {Swinbank}, {Szalay}, {Takacs},
  {Tether}, {Thaler}, {Thayer}, {Thomas}, {Thornton}, {Thukral}, {Tice},
  {Trilling}, {Turri}, {Van Berg}, {Vanden Berk}, {Vetter}, {Virieux},
  {Vucina}, {Wahl}, {Walkowicz}, {Walsh}, {Walter}, {Wang}, {Wang}, {Warner},
  {Wiecha}, {Willman}, {Winters}, {Wittman}, {Wolff}, {Wood-Vasey}, {Wu},
  {Xin}, {Yoachim}, \& {Zhan}}]{2019ApJ...873..111I}
{Ivezi{\'c}}, {\v{Z}}., {Kahn}, S.~M., {Tyson}, J.~A., {et~al.} 2019, \apj,
  873, 111

\bibitem[{{Jing} \& {Suto}(2000)}]{2000ApJ...529L..69J}
{Jing}, Y.~P. \& {Suto}, Y. 2000, \apjl, 529, L69

\bibitem[{{Joudaki} {et~al.}(2018){Joudaki}, {Blake}, {Johnson}, {Amon},
  {Asgari}, {Choi}, {Erben}, {Glazebrook}, {Harnois-D{\'e}raps}, {Heymans},
  {Hildebrandt}, {Hoekstra}, {Klaes}, {Kuijken}, {Lidman}, {Mead}, {Miller},
  {Parkinson}, {Poole}, {Schneider}, {Viola}, \& {Wolf}}]{2018MNRAS.474.4894J}
{Joudaki}, S., {Blake}, C., {Johnson}, A., {et~al.} 2018, \mnras, 474, 4894

\bibitem[{{Jullo} \& {Kneib}(2009)}]{2009MNRAS.395.1319J}
{Jullo}, E. \& {Kneib}, J.~P. 2009, \mnras, 395, 1319

\bibitem[{{Jullo} {et~al.}(2007){Jullo}, {Kneib}, {Limousin},
  {El{\'\i}asd{\'o}ttir}, {Marshall}, \& {Verdugo}}]{2007NJPh....9..447J}
{Jullo}, E., {Kneib}, J.~P., {Limousin}, M., {et~al.} 2007, New Journal of
  Physics, 9, 447

\bibitem[{{Jullo} {et~al.}(2010){Jullo}, {Natarajan}, {Kneib}, {D'Aloisio},
  {Limousin}, {Richard}, \& {Schimd}}]{2010Sci...329..924J}
{Jullo}, E., {Natarajan}, P., {Kneib}, J.~P., {et~al.} 2010, Science, 329, 924

\bibitem[{{Karman} {et~al.}(2017){Karman}, {Caputi}, {Caminha}, {Gronke},
  {Grillo}, {Balestra}, {Rosati}, {Vanzella}, {Coe}, {Dijkstra}, {Koekemoer},
  {McLeod}, {Mercurio}, \& {Nonino}}]{2017A&A...599A..28K}
{Karman}, W., {Caputi}, K.~I., {Caminha}, G.~B., {et~al.} 2017, \aap, 599, A28

\bibitem[{{Kassiola} \& {Kovner}(1993)}]{1993ApJ...417..450K}
{Kassiola}, A. \& {Kovner}, I. 1993, \apj, 417, 450

\bibitem[{{Kneib} {et~al.}(1996){Kneib}, {Ellis}, {Smail}, {Couch}, \&
  {Sharples}}]{1996ApJ...471..643K}
{Kneib}, J.~P., {Ellis}, R.~S., {Smail}, I., {Couch}, W.~J., \& {Sharples},
  R.~M. 1996, \apj, 471, 643

\bibitem[{{Kneib} \& {Natarajan}(2011)}]{2011A&ARv..19...47K}
{Kneib}, J.-P. \& {Natarajan}, P. 2011, \aapr, 19, 47

\bibitem[{{Komatsu} {et~al.}(2011){Komatsu}, {Smith}, {Dunkley}, {Bennett},
  {Gold}, {Hinshaw}, {Jarosik}, {Larson}, {Nolta}, {Page}, {Spergel},
  {Halpern}, {Hill}, {Kogut}, {Limon}, {Meyer}, {Odegard}, {Tucker}, {Weiland},
  {Wollack}, \& {Wright}}]{2011ApJS..192...18K}
{Komatsu}, E., {Smith}, K.~M., {Dunkley}, J., {et~al.} 2011, \apjs, 192, 18

\bibitem[{{Kowalski} {et~al.}(2008){Kowalski}, {Rubin}, {Aldering},
  {Agostinho}, {Amadon}, {Amanullah}, {Balland}, {Barbary}, {Blanc}, {Challis},
  {Conley}, {Connolly}, {Covarrubias}, {Dawson}, {Deustua}, {Ellis}, {Fabbro},
  {Fadeyev}, {Fan}, {Farris}, {Folatelli}, {Frye}, {Garavini}, {Gates},
  {Germany}, {Goldhaber}, {Goldman}, {Goobar}, {Groom}, {Haissinski}, {Hardin},
  {Hook}, {Kent}, {Kim}, {Knop}, {Lidman}, {Linder}, {Mendez}, {Meyers},
  {Miller}, {Moniez}, {Mour{\~a}o}, {Newberg}, {Nobili}, {Nugent}, {Pain},
  {Perdereau}, {Perlmutter}, {Phillips}, {Prasad}, {Quimby}, {Regnault},
  {Rich}, {Rubenstein}, {Ruiz-Lapuente}, {Santos}, {Schaefer}, {Schommer},
  {Smith}, {Soderberg}, {Spadafora}, {Strolger}, {Strovink}, {Suntzeff},
  {Suzuki}, {Thomas}, {Walton}, {Wang}, {Wood-Vasey}, \&
  {Yun}}]{2008ApJ...686..749K}
{Kowalski}, M., {Rubin}, D., {Aldering}, G., {et~al.} 2008, \apj, 686, 749

\bibitem[{{Lagattuta} {et~al.}(2019){Lagattuta}, {Richard}, {Bauer},
  {Cl{\'e}ment}, {Mahler}, {Soucail}, {Carton}, {Kneib}, {Laporte}, {Martinez},
  {Patr{\'\i}cio}, {Payne}, {Pell{\'o}}, {Schmidt}, \& {de la
  Vieuville}}]{2019MNRAS.485.3738L}
{Lagattuta}, D.~J., {Richard}, J., {Bauer}, F.~E., {et~al.} 2019, \mnras, 485,
  3738

\bibitem[{{Laureijs} {et~al.}(2011){Laureijs}, {Amiaux}, {Arduini},
  {Augu{\`e}res}, {Brinchmann}, {Cole}, {Cropper}, {Dabin}, {Duvet}, {Ealet},
  {Garilli}, {Gondoin}, {Guzzo}, {Hoar}, {Hoekstra}, {Holmes}, {Kitching},
  {Maciaszek}, {Mellier}, {Pasian}, {Percival}, {Rhodes}, {Saavedra Criado},
  {Sauvage}, {Scaramella}, {Valenziano}, {Warren}, {Bender}, {Castander},
  {Cimatti}, {Le F{\`e}vre}, {Kurki-Suonio}, {Levi}, {Lilje}, {Meylan},
  {Nichol}, {Pedersen}, {Popa}, {Rebolo Lopez}, {Rix}, {Rottgering},
  {Zeilinger}, {Grupp}, {Hudelot}, {Massey}, {Meneghetti}, {Miller}, {Paltani},
  {Paulin-Henriksson}, {Pires}, {Saxton}, {Schrabback}, {Seidel}, {Walsh},
  {Aghanim}, {Amendola}, {Bartlett}, {Baccigalupi}, {Beaulieu}, {Benabed},
  {Cuby}, {Elbaz}, {Fosalba}, {Gavazzi}, {Helmi}, {Hook}, {Irwin}, {Kneib},
  {Kunz}, {Mannucci}, {Moscardini}, {Tao}, {Teyssier}, {Weller}, {Zamorani},
  {Zapatero Osorio}, {Boulade}, {Foumond}, {Di Giorgio}, {Guttridge}, {James},
  {Kemp}, {Martignac}, {Spencer}, {Walton}, {Bl{\"u}mchen}, {Bonoli},
  {Bortoletto}, {Cerna}, {Corcione}, {Fabron}, {Jahnke}, {Ligori}, {Madrid},
  {Martin}, {Morgante}, {Pamplona}, {Prieto}, {Riva}, {Toledo}, {Trifoglio},
  {Zerbi}, {Abdalla}, {Douspis}, {Grenet}, {Borgani}, {Bouwens}, {Courbin},
  {Delouis}, {Dubath}, {Fontana}, {Frailis}, {Grazian}, {Koppenh{\"o}fer},
  {Mansutti}, {Melchior}, {Mignoli}, {Mohr}, {Neissner}, {Noddle}, {Poncet},
  {Scodeggio}, {Serrano}, {Shane}, {Starck}, {Surace}, {Taylor},
  {Verdoes-Kleijn}, {Vuerli}, {Williams}, {Zacchei}, {Altieri}, {Escudero
  Sanz}, {Kohley}, {Oosterbroek}, {Astier}, {Bacon}, {Bardelli}, {Baugh},
  {Bellagamba}, {Benoist}, {Bianchi}, {Biviano}, {Branchini}, {Carbone},
  {Cardone}, {Clements}, {Colombi}, {Conselice}, {Cresci}, {Deacon}, {Dunlop},
  {Fedeli}, {Fontanot}, {Franzetti}, {Giocoli}, {Garcia-Bellido}, {Gow},
  {Heavens}, {Hewett}, {Heymans}, {Holland}, {Huang}, {Ilbert}, {Joachimi},
  {Jennins}, {Kerins}, {Kiessling}, {Kirk}, {Kotak}, {Krause}, {Lahav}, {van
  Leeuwen}, {Lesgourgues}, {Lombardi}, {Magliocchetti}, {Maguire}, {Majerotto},
  {Maoli}, {Marulli}, {Maurogordato}, {McCracken}, {McLure}, {Melchiorri},
  {Merson}, {Moresco}, {Nonino}, {Norberg}, {Peacock}, {Pello}, {Penny},
  {Pettorino}, {Di Porto}, {Pozzetti}, {Quercellini}, {Radovich}, {Rassat},
  {Roche}, {Ronayette}, {Rossetti}, {Sartoris}, {Schneider}, {Semboloni},
  {Serjeant}, {Simpson}, {Skordis}, {Smadja}, {Smartt}, {Spano}, {Spiro},
  {Sullivan}, {Tilquin}, {Trotta}, {Verde}, {Wang}, {Williger}, {Zhao},
  {Zoubian}, \& {Zucca}}]{2011arXiv1110.3193L}
{Laureijs}, R., {Amiaux}, J., {Arduini}, S., {et~al.} 2011, arXiv e-prints,
  arXiv:1110.3193

\bibitem[{{Lewis}(2019)}]{2019arXiv191013970L}
{Lewis}, A. 2019, arXiv e-prints, arXiv:1910.13970

\bibitem[{{Linder}(2003)}]{2003PhRvL..90i1301L}
{Linder}, E.~V. 2003, \prl, 90, 091301

\bibitem[{{Link} \& {Pierce}(1998)}]{1998ApJ...502...63L}
{Link}, R. \& {Pierce}, M.~J. 1998, \apj, 502, 63

\bibitem[{{Lotz} {et~al.}(2017){Lotz}, {Koekemoer}, {Coe}, {Grogin}, {Capak},
  {Mack}, {Anderson}, {Avila}, {Barker}, {Borncamp}, {Brammer}, {Durbin},
  {Gunning}, {Hilbert}, {Jenkner}, {Khandrika}, {Levay}, {Lucas}, {MacKenty},
  {Ogaz}, {Porterfield}, {Reid}, {Robberto}, {Royle}, {Smith},
  {Storrie-Lombardi}, {Sunnquist}, {Surace}, {Taylor}, {Williams}, {Bullock},
  {Dickinson}, {Finkelstein}, {Natarajan}, {Richard}, {Robertson}, {Tumlinson},
  {Zitrin}, {Flanagan}, {Sembach}, {Soifer}, \&
  {Mountain}}]{2017ApJ...837...97L}
{Lotz}, J.~M., {Koekemoer}, A., {Coe}, D., {et~al.} 2017, \apj, 837, 97

\bibitem[{{Maga{\~n}a} {et~al.}(2018){Maga{\~n}a}, {Acebr{\'o}n}, {Motta},
  {Verdugo}, {Jullo}, \& {Limousin}}]{2018ApJ...865..122M}
{Maga{\~n}a}, J., {Acebr{\'o}n}, A., {Motta}, V., {et~al.} 2018, \apj, 865, 122

\bibitem[{{Maga{\~n}a} {et~al.}(2015){Maga{\~n}a}, {Motta}, {C{\'a}rdenas},
  {Verdugo}, \& {Jullo}}]{2015ApJ...813...69M}
{Maga{\~n}a}, J., {Motta}, V., {C{\'a}rdenas}, V.~H., {Verdugo}, T., \&
  {Jullo}, E. 2015, \apj, 813, 69

\bibitem[{{Mahler} {et~al.}(2018){Mahler}, {Richard}, {Cl{\'e}ment},
  {Lagattuta}, {Schmidt}, {Patr{\'\i}cio}, {Soucail}, {Bacon}, {Pello},
  {Bouwens}, {Maseda}, {Martinez}, {Carollo}, {Inami}, {Leclercq}, \&
  {Wisotzki}}]{2018MNRAS.473..663M}
{Mahler}, G., {Richard}, J., {Cl{\'e}ment}, B., {et~al.} 2018, \mnras, 473, 663

\bibitem[{{Martizzi} {et~al.}(2012){Martizzi}, {Teyssier}, {Moore}, \&
  {Wentz}}]{2012MNRAS.422.3081M}
{Martizzi}, D., {Teyssier}, R., {Moore}, B., \& {Wentz}, T. 2012, \mnras, 422,
  3081

\bibitem[{{Meneghetti} {et~al.}(2020){Meneghetti}, {Davoli}, {Bergamini},
  {Rosati}, {Natarajan}, {Giocoli}, {Caminha}, {Metcalf}, {Rasia}, {Borgani},
  {Calura}, {Grillo}, {Mercurio}, \& {Vanzella}}]{2020Sci...369.1347M}
{Meneghetti}, M., {Davoli}, G., {Bergamini}, P., {et~al.} 2020, Science, 369,
  1347

\bibitem[{{Millon} {et~al.}(2020{\natexlab{a}}){Millon}, {Courbin}, {Bonvin},
  {Buckley-Geer}, {Fassnacht}, {Frieman}, {Marshall}, {Suyu}, {Treu},
  {Anguita}, {Motta}, {Agnello}, {Chan}, {Chao}, {Chijani}, {Gilman},
  {Gilmore}, {Lemon}, {Lucey}, {Melo}, {Paic}, {Rojas}, {Sluse}, {Williams},
  {Hempel}, {Kim}, {Lachaume}, \& {Rabus}}]{2020A&A...642A.193M}
{Millon}, M., {Courbin}, F., {Bonvin}, V., {et~al.} 2020{\natexlab{a}}, \aap,
  642, A193

\bibitem[{{Millon} {et~al.}(2020{\natexlab{b}}){Millon}, {Courbin}, {Bonvin},
  {Paic}, {Meylan}, {Tewes}, {Sluse}, {Magain}, {Chan}, {Galan}, {Joseph},
  {Lemon}, {Tihhonova}, {Anderson}, {Marmier}, {Chazelas}, {Lendl}, {Triaud},
  \& {Wyttenbach}}]{2020A&A...640A.105M}
{Millon}, M., {Courbin}, F., {Bonvin}, V., {et~al.} 2020{\natexlab{b}}, \aap,
  640, A105

\bibitem[{{Monna} {et~al.}(2017){Monna}, {Seitz}, {Balestra}, {Rosati},
  {Grillo}, {Halkola}, {Suyu}, {Coe}, {Caminha}, {Frye}, {Koekemoer},
  {Mercurio}, {Nonino}, {Postman}, \& {Zitrin}}]{2017MNRAS.466.4094M}
{Monna}, A., {Seitz}, S., {Balestra}, I., {et~al.} 2017, \mnras, 466, 4094

\bibitem[{{Motta} {et~al.}(2021){Motta}, {Garc{\'\i}a-Aspeitia},
  {Hern{\'a}ndez-Almada}, {Maga{\~n}a}, \& {Verdugo}}]{2021Univ....7..163M}
{Motta}, V., {Garc{\'\i}a-Aspeitia}, M.~A., {Hern{\'a}ndez-Almada}, A.,
  {Maga{\~n}a}, J., \& {Verdugo}, T. 2021, Universe, 7, 163

\bibitem[{{Navarro} {et~al.}(1996){Navarro}, {Frenk}, \&
  {White}}]{1996ApJ...462..563N}
{Navarro}, J.~F., {Frenk}, C.~S., \& {White}, S.~D.~M. 1996, \apj, 462, 563

\bibitem[{{Navarro} {et~al.}(1997){Navarro}, {Frenk}, \&
  {White}}]{1997ApJ...490..493N}
{Navarro}, J.~F., {Frenk}, C.~S., \& {White}, S.~D.~M. 1997, \apj, 490, 493

\bibitem[{{Newman} {et~al.}(2011){Newman}, {Treu}, {Ellis}, \&
  {Sand}}]{2011ApJ...728L..39N}
{Newman}, A.~B., {Treu}, T., {Ellis}, R.~S., \& {Sand}, D.~J. 2011, \apjl, 728,
  L39

\bibitem[{{Newman} {et~al.}(2013{\natexlab{a}}){Newman}, {Treu}, {Ellis}, \&
  {Sand}}]{2013ApJ...765...25N}
{Newman}, A.~B., {Treu}, T., {Ellis}, R.~S., \& {Sand}, D.~J.
  2013{\natexlab{a}}, \apj, 765, 25

\bibitem[{{Newman} {et~al.}(2013{\natexlab{b}}){Newman}, {Treu}, {Ellis},
  {Sand}, {Nipoti}, {Richard}, \& {Jullo}}]{2013ApJ...765...24N}
{Newman}, A.~B., {Treu}, T., {Ellis}, R.~S., {et~al.} 2013{\natexlab{b}}, \apj,
  765, 24

\bibitem[{{Perlmutter} {et~al.}(1999){Perlmutter}, {Aldering}, {Goldhaber},
  {Knop}, {Nugent}, {Castro}, {Deustua}, {Fabbro}, {Goobar}, {Groom}, {Hook},
  {Kim}, {Kim}, {Lee}, {Nunes}, {Pain}, {Pennypacker}, {Quimby}, {Lidman},
  {Ellis}, {Irwin}, {McMahon}, {Ruiz-Lapuente}, {Walton}, {Schaefer}, {Boyle},
  {Filippenko}, {Matheson}, {Fruchter}, {Panagia}, {Newberg}, {Couch}, \&
  {Project}}]{1999ApJ...517..565P}
{Perlmutter}, S., {Aldering}, G., {Goldhaber}, G., {et~al.} 1999, \apj, 517,
  565

\bibitem[{{Planck Collaboration} {et~al.}(2020){Planck Collaboration},
  {Aghanim}, {Akrami}, {Ashdown}, {Aumont}, {Baccigalupi}, {Ballardini},
  {Banday}, {Barreiro}, {Bartolo}, {Basak}, {Battye}, {Benabed}, {Bernard},
  {Bersanelli}, {Bielewicz}, {Bock}, {Bond}, {Borrill}, {Bouchet}, {Boulanger},
  {Bucher}, {Burigana}, {Butler}, {Calabrese}, {Cardoso}, {Carron},
  {Challinor}, {Chiang}, {Chluba}, {Colombo}, {Combet}, {Contreras}, {Crill},
  {Cuttaia}, {de Bernardis}, {de Zotti}, {Delabrouille}, {Delouis}, {Di
  Valentino}, {Diego}, {Dor{\'e}}, {Douspis}, {Ducout}, {Dupac}, {Dusini},
  {Efstathiou}, {Elsner}, {En{\ss}lin}, {Eriksen}, {Fantaye}, {Farhang},
  {Fergusson}, {Fernandez-Cobos}, {Finelli}, {Forastieri}, {Frailis},
  {Fraisse}, {Franceschi}, {Frolov}, {Galeotta}, {Galli}, {Ganga},
  {G{\'e}nova-Santos}, {Gerbino}, {Ghosh}, {Gonz{\'a}lez-Nuevo}, {G{\'o}rski},
  {Gratton}, {Gruppuso}, {Gudmundsson}, {Hamann}, {Handley}, {Hansen},
  {Herranz}, {Hildebrandt}, {Hivon}, {Huang}, {Jaffe}, {Jones}, {Karakci},
  {Keih{\"a}nen}, {Keskitalo}, {Kiiveri}, {Kim}, {Kisner}, {Knox},
  {Krachmalnicoff}, {Kunz}, {Kurki-Suonio}, {Lagache}, {Lamarre}, {Lasenby},
  {Lattanzi}, {Lawrence}, {Le Jeune}, {Lemos}, {Lesgourgues}, {Levrier},
  {Lewis}, {Liguori}, {Lilje}, {Lilley}, {Lindholm}, {L{\'o}pez-Caniego},
  {Lubin}, {Ma}, {Mac{\'\i}as-P{\'e}rez}, {Maggio}, {Maino}, {Mandolesi},
  {Mangilli}, {Marcos-Caballero}, {Maris}, {Martin}, {Martinelli},
  {Mart{\'\i}nez-Gonz{\'a}lez}, {Matarrese}, {Mauri}, {McEwen}, {Meinhold},
  {Melchiorri}, {Mennella}, {Migliaccio}, {Millea}, {Mitra},
  {Miville-Desch{\^e}nes}, {Molinari}, {Montier}, {Morgante}, {Moss}, {Natoli},
  {N{\o}rgaard-Nielsen}, {Pagano}, {Paoletti}, {Partridge}, {Patanchon},
  {Peiris}, {Perrotta}, {Pettorino}, {Piacentini}, {Polastri}, {Polenta},
  {Puget}, {Rachen}, {Reinecke}, {Remazeilles}, {Renzi}, {Rocha}, {Rosset},
  {Roudier}, {Rubi{\~n}o-Mart{\'\i}n}, {Ruiz-Granados}, {Salvati}, {Sandri},
  {Savelainen}, {Scott}, {Shellard}, {Sirignano}, {Sirri}, {Spencer},
  {Sunyaev}, {Suur-Uski}, {Tauber}, {Tavagnacco}, {Tenti}, {Toffolatti},
  {Tomasi}, {Trombetti}, {Valenziano}, {Valiviita}, {Van Tent}, {Vibert},
  {Vielva}, {Villa}, {Vittorio}, {Wandelt}, {Wehus}, {White}, {White},
  {Zacchei}, \& {Zonca}}]{2020A&A...641A...6P}
{Planck Collaboration}, {Aghanim}, N., {Akrami}, Y., {et~al.} 2020, \aap, 641,
  A6

\bibitem[{{Porredon} {et~al.}(2021){Porredon}, {Crocce}, {Elvin-Poole},
  {Cawthon}, {Giannini}, {De Vicente}, {Carnero Rosell}, {Ferrero}, {Krause},
  {Fang}, {Prat}, {Rodriguez-Monroy}, {Pandey}, {Pocino}, {Castander}, {Choi},
  {Amon}, {Tutusaus}, {Dodelson}, {Sevilla-Noarbe}, {Fosalba}, {Gaztanaga},
  {Alarcon}, {Alves}, {Andrade-Oliveira}, {Baxter}, {Bechtol}, {Becker},
  {Bernstein}, {Blazek}, {Camacho}, {Campos}, {Carrasco Kind}, {Chintalapati},
  {Cordero}, {DeRose}, {Di Valentino}, {Doux}, {Eifler}, {Everett},
  {Fert{\'e}}, {Friedrich}, {Gatti}, {Gruen}, {Harrison}, {Hartley}, {Herner},
  {Huff}, {Huterer}, {Jain}, {Jarvis}, {Lee}, {Lemos}, {MacCrann},
  {Mena-Fern{\'a}ndez}, {Muir}, {Myles}, {Park}, {Raveri}, {Rosenfeld}, {Ross},
  {Rykoff}, {Samuroff}, {S{\'a}nchez}, {Sanchez}, {Sanchez}, {Sanchez Cid},
  {Scolnic}, {Secco}, {Sheldon}, {Troja}, {Troxel}, {Weaverdyck}, {Yanny},
  {Zuntz}, {Abbott}, {Aguena}, {Allam}, {Annis}, {Avila}, {Bacon}, {Bertin},
  {Bhargava}, {Brooks}, {Buckley-Geer}, {Burke}, {Carretero}, {Costanzi}, {da
  Costa}, {Pereira}, {Davis}, {Desai}, {Diehl}, {Dietrich}, {Doel},
  {Drlica-Wagner}, {Eckert}, {Evrard}, {Flaugher}, {Frieman},
  {Garc{\'\i}a-Bellido}, {Gerdes}, {Giannantonio}, {Gruendl}, {Gschwend},
  {Gutierrez}, {Hinton}, {Hollowood}, {Honscheid}, {Hoyle}, {James}, {Kuehn},
  {Kuropatkin}, {Lahav}, {Lidman}, {Lima}, {Lin}, {Maia}, {Marshall},
  {Martini}, {Melchior}, {Menanteau}, {Miquel}, {Mohr}, {Morgan}, {Ogando},
  {Palmese}, {Paz-Chinch{\'o}n}, {Petravick}, {Pieres}, {Plazas Malag{\'o}n},
  {Romer}, {Santiago}, {Scarpine}, {Schubnell}, {Serrano}, {Smith},
  {Soares-Santos}, {Suchyta}, {Tarle}, {Thomas}, {To}, {Varga}, \&
  {Weller}}]{2021arXiv210513546P}
{Porredon}, A., {Crocce}, M., {Elvin-Poole}, J., {et~al.} 2021, arXiv e-prints,
  arXiv:2105.13546

\bibitem[{{Postman} {et~al.}(2012){Postman}, {Coe}, {Ben{\'\i}tez}, {Bradley},
  {Broadhurst}, {Donahue}, {Ford}, {Graur}, {Graves}, {Jouvel}, {Koekemoer},
  {Lemze}, {Medezinski}, {Molino}, {Moustakas}, {Ogaz}, {Riess}, {Rodney},
  {Rosati}, {Umetsu}, {Zheng}, {Zitrin}, {Bartelmann}, {Bouwens}, {Czakon},
  {Golwala}, {Host}, {Infante}, {Jha}, {Jimenez-Teja}, {Kelson}, {Lahav},
  {Lazkoz}, {Maoz}, {McCully}, {Melchior}, {Meneghetti}, {Merten}, {Moustakas},
  {Nonino}, {Patel}, {Reg{\"o}s}, {Sayers}, {Seitz}, \& {Van der
  Wel}}]{2012ApJS..199...25P}
{Postman}, M., {Coe}, D., {Ben{\'\i}tez}, N., {et~al.} 2012, \apjs, 199, 25

\bibitem[{{Richard} {et~al.}(2021){Richard}, {Claeyssens}, {Lagattuta},
  {Guaita}, {Bauer}, {Pello}, {Carton}, {Bacon}, {Soucail}, {Lyon}, {Kneib},
  {Mahler}, {Cl{\'e}ment}, {Mercier}, {Variu}, {Tamone}, {Ebeling}, {Schmidt},
  {Nanayakkara}, {Maseda}, {Weilbacher}, {Bouch{\'e}}, {Bouwens}, {Wisotzki},
  {de la Vieuville}, {Martinez}, \& {Patr{\'\i}cio}}]{2021A&A...646A..83R}
{Richard}, J., {Claeyssens}, A., {Lagattuta}, D., {et~al.} 2021, \aap, 646, A83

\bibitem[{{Richard} {et~al.}(2015){Richard}, {Patricio}, {Martinez}, {Bacon},
  {Clement}, {Weilbacher}, {Soto}, {Wisotzki}, {Vernet}, {Pello}, {Schaye},
  {Turner}, \& {Martinsson}}]{2015MNRAS.446L..16R}
{Richard}, J., {Patricio}, V., {Martinez}, J., {et~al.} 2015, \mnras, 446, L16

\bibitem[{{Riess} {et~al.}(1998){Riess}, {Filippenko}, {Challis},
  {Clocchiatti}, {Diercks}, {Garnavich}, {Gilliland}, {Hogan}, {Jha},
  {Kirshner}, {Leibundgut}, {Phillips}, {Reiss}, {Schmidt}, {Schommer},
  {Smith}, {Spyromilio}, {Stubbs}, {Suntzeff}, \&
  {Tonry}}]{1998AJ....116.1009R}
{Riess}, A.~G., {Filippenko}, A.~V., {Challis}, P., {et~al.} 1998, \aj, 116,
  1009

\bibitem[{{Rosati} {et~al.}(2014){Rosati}, {Balestra}, {Grillo}, {Mercurio},
  {Nonino}, {Biviano}, {Girardi}, {Vanzella}, \& {Clash-VLT
  Team}}]{2014Msngr.158...48R}
{Rosati}, P., {Balestra}, I., {Grillo}, C., {et~al.} 2014, The Messenger, 158,
  48

\bibitem[{{Ross} {et~al.}(2017){Ross}, {Beutler}, {Chuang}, {Pellejero-Ibanez},
  {Seo}, {Vargas-Maga{\~n}a}, {Cuesta}, {Percival}, {Burden}, {S{\'a}nchez},
  {Grieb}, {Reid}, {Brownstein}, {Dawson}, {Eisenstein}, {Ho}, {Kitaura},
  {Nichol}, {Olmstead}, {Prada}, {Rodr{\'\i}guez-Torres}, {Saito},
  {Salazar-Albornoz}, {Schneider}, {Thomas}, {Tinker}, {Tojeiro}, {Wang},
  {White}, \& {Zhao}}]{2017MNRAS.464.1168R}
{Ross}, A.~J., {Beutler}, F., {Chuang}, C.-H., {et~al.} 2017, \mnras, 464, 1168

\bibitem[{{Sand} {et~al.}(2004){Sand}, {Treu}, {Smith}, \&
  {Ellis}}]{2004ApJ...604...88S}
{Sand}, D.~J., {Treu}, T., {Smith}, G.~P., \& {Ellis}, R.~S. 2004, \apj, 604,
  88

\bibitem[{{Scaramella} {et~al.}(2021){Scaramella}, {Amiaux}, {Mellier},
  {Burigana}, {Carvalho}, {Cuillandre}, {Da Silva}, {Derosa}, {Dinis},
  {Maiorano}, {Maris}, {Tereno}, {Laureijs}, {Boenke}, {Buenadicha}, {Dupac},
  {Gaspar Venancio}, {G{\'o}mez-{\'A}lvarez}, {Hoar}, {Alvarez}, {Racca},
  {Saavedra-Criado}, {Schwartz}, {Vavrek}, {Schirmer}, {Aussel}, {Azzollini},
  {Cardone}, {Cropper}, {Ealet}, {Garilli}, {Gillard}, {Granett}, {Guzzo},
  {Hoekstra}, {Jahnke}, {Kitching}, {Meneghetti}, {Miller}, {Nakajima},
  {Niemi}, {Pasian}, {Percival}, {Sauvage}, {Scodeggio}, {Wachter}, {Zacchei},
  {Aghanim}, {Amara}, {Auphan}, {Auricchio}, {Awan}, {Balestra}, {Bender},
  {Bodendorf}, {Bonino}, {Branchini}, {Brau-Nogue}, {Brescia}, {Candini},
  {Capobianco}, {Carbone}, {Carlberg}, {Carretero}, {Casas}, {Castander},
  {Castellano}, {Cavuoti}, {Cimatti}, {Cledassou}, {Congedo}, {Conselice},
  {Conversi}, {Copin}, {Corcione}, {Costille}, {Courbin}, {Degaudenzi},
  {Douspis}, {Dubath}, {Duncan}, {Dusini}, {Farrens}, {Ferriol}, {Fosalba},
  {Fourmanoit}, {Frailis}, {Franceschi}, {Franzetti}, {Fumana}, {Gillis},
  {Giocoli}, {Grazian}, {Grupp}, {Haugan}, {Holmes}, {Hormuth}, {Hudelot},
  {Kermiche}, {Kiessling}, {Kilbinger}, {Kohley}, {Kubik}, {K{\"u}mmel},
  {Kunz}, {Kurki-Suonio}, {Ligori}, {Lilje}, {Lloro}, {Mansutti}, {Marggraf},
  {Markovic}, {Marulli}, {Massey}, {Maurogordato}, {Melchior}, {Merlin},
  {Meylan}, {Mohr}, {Moresco}, {Morin}, {Moscardini}, {Munari}, {Nichol},
  {Padilla}, {Paltani}, {Peacock}, {Pedersen}, {Pettorino}, {Pires}, {Poncet},
  {Popa}, {Pozzetti}, {Raison}, {Rebolo}, {Rhodes}, {Rix}, {Roncarelli},
  {Rossetti}, {Saglia}, {Schneider}, {Schrabback}, {Secroun}, {Seidel},
  {Serrano}, {Sirignano}, {Sirri}, {Skottfelt}, {Stanco}, {Starck},
  {Tallada-Cresp{\'\i}}, {Tavagnacco}, {Taylor}, {Teplitz}, {Toledo-Moreo},
  {Torradeflot}, {Trifoglio}, {Valentijn}, {Valenziano}, {Verdoes Kleijn},
  {Wang}, {Welikala}, {Weller}, {Wetzstein}, {Zamorani}, {Zoubian}, {Andreon},
  {Baldi}, {Bardelli}, {Boucaud}, {Camera}, {Fabbian}, {Farinelli},
  {Graci{\'a}-Carpio}, {Maino}, {Medinaceli}, {Mei}, {Neissner}, {Polenta},
  {Renzi}, {Romelli}, {Rosset}, {Sureau}, {Tenti}, {Vassallo}, {Zucca},
  {Baccigalupi}, {Balaguera-Antol{\'\i}nez}, {Battaglia}, {Biviano}, {Borgani},
  {Bozzo}, {Cabanac}, {Cappi}, {Casas}, {Castignani}, {Colodro-Conde},
  {Coupon}, {Courtois}, {Cuby}, {de la Torre}, {Desai}, {Di Ferdinando},
  {Dole}, {Fabricius}, {Farina}, {Ferreira}, {Finelli}, {Flose-Reimberg},
  {Fotopoulou}, {Galeotta}, {Ganga}, {Gozaliasl}, {Hook}, {Keihanen},
  {Kirkpatrick}, {Liebing}, {Lindholm}, {Mainetti}, {Martinelli}, {Martinet},
  {Maturi}, {McCracken}, {Metcalf}, {Morgante}, {Nightingale}, {Nucita},
  {Patrizii}, {Potter}, {Riccio}, {S{\'a}nchez}, {Sapone}, {Schewtschenko},
  {Schultheis}, {Scottez}, {Teyssier}, {Tutusaus}, {Valiviita}, {Viel},
  {Vriend}, \& {Whittaker}}]{2021arXiv210801201S}
{Scaramella}, R., {Amiaux}, J., {Mellier}, Y., {et~al.} 2021, arXiv e-prints,
  arXiv:2108.01201

\bibitem[{{Schaller} {et~al.}(2015){Schaller}, {Frenk}, {Bower}, {Theuns},
  {Trayford}, {Crain}, {Furlong}, {Schaye}, {Dalla Vecchia}, \&
  {McCarthy}}]{2015MNRAS.452..343S}
{Schaller}, M., {Frenk}, C.~S., {Bower}, R.~G., {et~al.} 2015, \mnras, 452, 343

\bibitem[{{Schneider}(2014)}]{2014A&A...568L...2S}
{Schneider}, P. 2014, \aap, 568, L2

\bibitem[{{Schneider} {et~al.}(1992){Schneider}, {Ehlers}, \&
  {Falco}}]{1992grle.book.....S}
{Schneider}, P., {Ehlers}, J., \& {Falco}, E.~E. 1992, {Gravitational Lenses}

\bibitem[{Schwarz(1978)}]{schwarz1978}
Schwarz, G. 1978, Ann. Statist., 6, 461

\bibitem[{{Scolnic} {et~al.}(2018){Scolnic}, {Jones}, {Rest}, {Pan},
  {Chornock}, {Foley}, {Huber}, {Kessler}, {Narayan}, {Riess}, {Rodney},
  {Berger}, {Brout}, {Challis}, {Drout}, {Finkbeiner}, {Lunnan}, {Kirshner},
  {Sanders}, {Schlafly}, {Smartt}, {Stubbs}, {Tonry}, {Wood-Vasey}, {Foley},
  {Hand}, {Johnson}, {Burgett}, {Chambers}, {Draper}, {Hodapp}, {Kaiser},
  {Kudritzki}, {Magnier}, {Metcalfe}, {Bresolin}, {Gall}, {Kotak}, {McCrum}, \&
  {Smith}}]{2018ApJ...859..101S}
{Scolnic}, D.~M., {Jones}, D.~O., {Rest}, A., {et~al.} 2018, \apj, 859, 101

\bibitem[{{Shajib} {et~al.}(2020){Shajib}, {Birrer}, {Treu}, {Agnello},
  {Buckley-Geer}, {Chan}, {Christensen}, {Lemon}, {Lin}, {Millon}, {Poh},
  {Rusu}, {Sluse}, {Spiniello}, {Chen}, {Collett}, {Courbin}, {Fassnacht},
  {Frieman}, {Galan}, {Gilman}, {More}, {Anguita}, {Auger}, {Bonvin},
  {McMahon}, {Meylan}, {Wong}, {Abbott}, {Annis}, {Avila}, {Bechtol}, {Brooks},
  {Brout}, {Burke}, {Carnero Rosell}, {Carrasco Kind}, {Carretero},
  {Castander}, {Costanzi}, {da Costa}, {De Vicente}, {Desai}, {Dietrich},
  {Doel}, {Drlica-Wagner}, {Evrard}, {Finley}, {Flaugher}, {Fosalba},
  {Garc{\'\i}a-Bellido}, {Gerdes}, {Gruen}, {Gruendl}, {Gschwend}, {Gutierrez},
  {Hollowood}, {Honscheid}, {Huterer}, {James}, {Jeltema}, {Krause},
  {Kuropatkin}, {Li}, {Lima}, {MacCrann}, {Maia}, {Marshall}, {Melchior},
  {Miquel}, {Ogando}, {Palmese}, {Paz-Chinch{\'o}n}, {Plazas}, {Romer},
  {Roodman}, {Sako}, {Sanchez}, {Santiago}, {Scarpine}, {Schubnell}, {Scolnic},
  {Serrano}, {Sevilla-Noarbe}, {Smith}, {Soares-Santos}, {Suchyta}, {Tarle},
  {Thomas}, {Walker}, \& {Zhang}}]{2020MNRAS.494.6072S}
{Shajib}, A.~J., {Birrer}, S., {Treu}, T., {et~al.} 2020, \mnras, 494, 6072

\bibitem[{{Smith} \& {Collett}(2021)}]{2021MNRAS.505.2136S}
{Smith}, R.~J. \& {Collett}, T.~E. 2021, \mnras, 505, 2136

\bibitem[{{Smoot} {et~al.}(1992){Smoot}, {Bennett}, {Kogut}, {Wright}, {Aymon},
  {Boggess}, {Cheng}, {de Amici}, {Gulkis}, {Hauser}, {Hinshaw}, {Jackson},
  {Janssen}, {Kaita}, {Kelsall}, {Keegstra}, {Lineweaver}, {Loewenstein},
  {Lubin}, {Mather}, {Meyer}, {Moseley}, {Murdock}, {Rokke}, {Silverberg},
  {Tenorio}, {Weiss}, \& {Wilkinson}}]{1992ApJ...396L...1S}
{Smoot}, G.~F., {Bennett}, C.~L., {Kogut}, A., {et~al.} 1992, \apjl, 396, L1

\bibitem[{{Spergel} {et~al.}(2015){Spergel}, {Gehrels}, {Baltay}, {Bennett},
  {Breckinridge}, {Donahue}, {Dressler}, {Gaudi}, {Greene}, {Guyon}, {Hirata},
  {Kalirai}, {Kasdin}, {Macintosh}, {Moos}, {Perlmutter}, {Postman},
  {Rauscher}, {Rhodes}, {Wang}, {Weinberg}, {Benford}, {Hudson}, {Jeong},
  {Mellier}, {Traub}, {Yamada}, {Capak}, {Colbert}, {Masters}, {Penny},
  {Savransky}, {Stern}, {Zimmerman}, {Barry}, {Bartusek}, {Carpenter}, {Cheng},
  {Content}, {Dekens}, {Demers}, {Grady}, {Jackson}, {Kuan}, {Kruk}, {Melton},
  {Nemati}, {Parvin}, {Poberezhskiy}, {Peddie}, {Ruffa}, {Wallace}, {Whipple},
  {Wollack}, \& {Zhao}}]{2015arXiv150303757S}
{Spergel}, D., {Gehrels}, N., {Baltay}, C., {et~al.} 2015, arXiv e-prints,
  arXiv:1503.03757

\bibitem[{{Suyu} \& {Halkola}(2010)}]{2010A&A...524A..94S}
{Suyu}, S.~H. \& {Halkola}, A. 2010, \aap, 524, A94

\bibitem[{{Suyu} {et~al.}(2010){Suyu}, {Marshall}, {Auger}, {Hilbert},
  {Blandford}, {Koopmans}, {Fassnacht}, \& {Treu}}]{2010ApJ...711..201S}
{Suyu}, S.~H., {Marshall}, P.~J., {Auger}, M.~W., {et~al.} 2010, \apj, 711, 201

\bibitem[{{van Uitert} {et~al.}(2018){van Uitert}, {Joachimi}, {Joudaki},
  {Amon}, {Heymans}, {K{\"o}hlinger}, {Asgari}, {Blake}, {Choi}, {Erben},
  {Farrow}, {Harnois-D{\'e}raps}, {Hildebrandt}, {Hoekstra}, {Kitching},
  {Klaes}, {Kuijken}, {Merten}, {Miller}, {Nakajima}, {Schneider}, {Valentijn},
  \& {Viola}}]{2018MNRAS.476.4662V}
{van Uitert}, E., {Joachimi}, B., {Joudaki}, S., {et~al.} 2018, \mnras, 476,
  4662

\bibitem[{{Vargas-Maga{\~n}a} {et~al.}(2018){Vargas-Maga{\~n}a}, {Ho},
  {Cuesta}, {O'Connell}, {Ross}, {Eisenstein}, {Percival}, {Grieb},
  {S{\'a}nchez}, {Tinker}, {Tojeiro}, {Beutler}, {Chuang}, {Kitaura}, {Prada},
  {Rodr{\'\i}guez-Torres}, {Rossi}, {Seo}, {Brownstein}, {Olmstead}, \&
  {Thomas}}]{2018MNRAS.477.1153V}
{Vargas-Maga{\~n}a}, M., {Ho}, S., {Cuesta}, A.~J., {et~al.} 2018, \mnras, 477,
  1153

\bibitem[{{Wong} {et~al.}(2020){Wong}, {Suyu}, {Chen}, {Rusu}, {Millon},
  {Sluse}, {Bonvin}, {Fassnacht}, {Taubenberger}, {Auger}, {Birrer}, {Chan},
  {Courbin}, {Hilbert}, {Tihhonova}, {Treu}, {Agnello}, {Ding}, {Jee},
  {Komatsu}, {Shajib}, {Sonnenfeld}, {Blandford}, {Koopmans}, {Marshall}, \&
  {Meylan}}]{2020MNRAS.498.1420W}
{Wong}, K.~C., {Suyu}, S.~H., {Chen}, G. C.~F., {et~al.} 2020, \mnras, 498,
  1420

\bibitem[{{Wright}(2007)}]{2007ApJ...664..633W}
{Wright}, E.~L. 2007, \apj, 664, 633

\bibitem[{{Wyithe} {et~al.}(2001){Wyithe}, {Turner}, \&
  {Spergel}}]{2001ApJ...555..504W}
{Wyithe}, J.~S.~B., {Turner}, E.~L., \& {Spergel}, D.~N. 2001, \apj, 555, 504

\bibitem[{{Zhao}(1996)}]{1996MNRAS.278..488Z}
{Zhao}, H. 1996, \mnras, 278, 488

\end{thebibliography}

\begin{appendix}

\section{NFW and gNFW models}
\label{ap:nfw_models}

In order to test the adopted mass profile parameterization, we have also considered the NFW and generalised-NFW \citep[gNFW,][]{1996MNRAS.278..488Z, 2000ApJ...529L..69J, 2001ApJ...555..504W} profiles to describe the cluster-scale component.
These parameterizations are less accurate in predicting the positions of the observed multiple images \citep[see e.g.][]{2010Sci...329..924J, 2015ApJ...800...38G, 2019A&A...632A..36C}.

The summaries with the best fit values for the NFW and gNFW models are shown in Tables \ref{tab:cosmo_models_nfw} and \ref{tab:cosmo_models_gnfw}, respectively.
In all cases, the NFW model provides a higher rms when comparing it to the corresponding model with a PIEMD profile (see Table \ref{tab:cosmo_models}).
The gNFW model provides marginally better rms values for the clusters MACS~J1931 and MACS~J0329 with an improvement not better than $\approx 15\%$.
However, the Bayesian information criteria \citep{schwarz1978} always increase (see column $\delta$BIC in table \ref{tab:cosmo_models_nfw}) when compared to the PIEMD models, thus the introduction of an additional free parameter in the gNFW models is not justified.
We note that the implementation of these models in the \emph{lenstool} software assumes elliptical symmetry in the lens potential due to numerical simplicity, instead of a more realistic elliptical mass distribution.
We thus adopt the PIEMD parameterization as reference in this work.

\begin{table}
\centering
\tiny
\caption{Summary of the NFW models.}
\begin{tabular}{l c c c c c c } \hline \hline
Model ID &  DOF & $\rm N_{free}$ & rms[\arcsec] & $\rm \chi^2/DOF$ & $\rm \delta BIC$\\
\hline
R2129 $\Omega_{\rm m}$-NFW                 & 21 &  9& 0.29 & 0.35 & 1.08 \\ 
R2129 $\Omega_{\rm m},w$-NFW               & 20 & 10& 0.29 & 0.36 & 1.07 \\ 
R2129 $\Omega_{\rm m},\Omega_{\rm k}$-NFW  & 20 & 10& 0.29 & 0.36 & 1.07 \\ 
R2129 $\Omega_{\rm m},w_0, w_{\rm a}$-NFW            & 19 & 11& 0.29 & 0.38 & 1.07 \\ 
\hline
A1063 $\Omega_{\rm m}$-NFW                 & 55 & 15& 0.38 & 0.58 & 1.02 \\ 
A1063 $\Omega_{\rm m},w$-NFW               & 54 & 16& 0.37 & 0.56 & 1.01 \\ 
A1063 $\Omega_{\rm m},\Omega_{\rm k}$-NFW  & 54 & 16& 0.38 & 0.58 & 1.01 \\ 
A1063 $\Omega_{\rm m},w_0, w_{\rm a}$-NFW            & 53 & 17& 0.36 & 0.54 & 1.00 \\ 
\hline
M1931 $\Omega_{\rm m}$-NFW                 & 11 & 13& 0.40 & 1.11 & 1.02 \\ 
M1931 $\Omega_{\rm m},w$-NFW               & 10 & 14& 0.39 & 1.18 & 1.02 \\ 
M1931 $\Omega_{\rm m},\Omega_{\rm k}$-NFW  & 10 & 14& 0.40 & 1.20 & 1.02 \\ 
M1931 $\Omega_{\rm m},w_0, w_{\rm a}$-NFW            &  9 & 15& 0.37 & 1.16 & 1.01 \\ 
\hline
M0329 $\Omega_{\rm m}$-NFW                 & 11 & 17& 0.27 & 0.60 & 1.02 \\ 
M0329 $\Omega_{\rm m},w$-NFW               & 10 & 18& 0.24 & 0.55 & 1.02 \\ 
M0329 $\Omega_{\rm m},\Omega_{\rm k}$-NFW  & 10 & 18& 0.27 & 0.65 & 1.02 \\ 
M0329 $\Omega_{\rm m},w_0, w_{\rm a}$-NFW            &  9 & 19& 0.24 & 0.60 & 1.02 \\ 
\hline
M2129 $\Omega_{\rm m}$-NFW                 & 39 & 15& 0.87 & 2.95 & 1.48 \\ 
M2129 $\Omega_{\rm m},w$-NFW               & 38 & 16& 0.86 & 2.92 & 1.46 \\ 
M2129 $\Omega_{\rm m},\Omega_{\rm k}$-NFW  & 38 & 16& 0.85 & 2.89 & 1.43 \\ 
M2129 $\Omega_{\rm m},w_0, w_{\rm a}$-NFW            & 37 & 17& 0.84 & 2.93 & 1.40 \\ 
\hline
\end{tabular}
\label{tab:cosmo_models_nfw}
\tablefoot{In addition to the quantities shown in Table \ref{tab:cosmo_models}, here we also list the ratio of the Bayesian Information Criteria between the NFW and the corresponding PIEMD models in the column $\rm \delta BIC$.}
\end{table}

\begin{table}
\centering
\tiny
\caption{Summary of the gNFW models.}
\begin{tabular}{l c c c c c c } \hline \hline
Model ID &  DOF & $\rm N_{free}$ & rms[\arcsec] & $\rm \chi^2/DOF$ & $\rm \delta BIC$\\
\hline
R2129 $\Omega_{\rm m}$-gNFW                & 20 & 10& 0.21 & 0.20 & 1.08 \\ 
R2129 $\Omega_{\rm m},w$-gNFW              & 19 & 11& 0.21 & 0.21 & 1.07 \\ 
R2129 $\Omega_{\rm m},\Omega_{\rm k}$-gNFW & 19 & 11& 0.21 & 0.21 & 1.07 \\ 
R2129 $\Omega_{\rm m},w_0, w_{\rm a}$-gNFW           & 18 & 12& 0.21 & 0.21 & 1.07 \\ 
\hline
A1063 $\Omega_{\rm m}$-gNFW                & 54 & 16& 0.37 & 0.57 & 1.04 \\ 
A1063 $\Omega_{\rm m},w$-gNFW              & 53 & 17& 0.36 & 0.53 & 1.02 \\ 
A1063 $\Omega_{\rm m},\Omega_{\rm k}$-gNFW & 53 & 17& 0.36 & 0.55 & 1.03 \\ 
A1063 $\Omega_{\rm m},w_0, w_{\rm a}$-gNFW           & 52 & 18& 0.36 & 0.53 & 1.02 \\ 
\hline
M1931 $\Omega_{\rm m}$-gNFW                & 10 & 14& 0.34 & 0.89 & 1.02 \\ 
M1931 $\Omega_{\rm m},w$-gNFW              &  9 & 15& 0.34 & 0.98 & 1.02 \\ 
M1931 $\Omega_{\rm m},\Omega_{\rm k}$-gNFW &  9 & 15& 0.34 & 0.99 & 1.02 \\ 
M1931 $\Omega_{\rm m},w_0, w_{\rm a}$-gNFW           &  8 & 16& 0.30 & 0.87 & 1.01 \\ 
\hline
M0329 $\Omega_{\rm m}$-gNFW                & 10 & 18& 0.21 & 0.41 & 1.03 \\ 
M0329 $\Omega_{\rm m},w$-gNFW              &  9 & 19& 0.20 & 0.42 & 1.04 \\ 
M0329 $\Omega_{\rm m},\Omega_{\rm k}$-gNFW &  9 & 19& 0.21 & 0.45 & 1.03 \\ 
M0329 $\Omega_{\rm m},w_0, w_{\rm a}$-gNFW           &  8 & 20& 0.20 & 0.47 & 1.04 \\ 
\hline
M2129 $\Omega_{\rm m}$-gNFW                & 38 & 16& 0.56 & 1.25 & 1.03 \\ 
M2129 $\Omega_{\rm m},w$-gNFW              & 37 & 17& 0.56 & 1.29 & 1.05 \\ 
M2129 $\Omega_{\rm m},\Omega_{\rm k}$-gNFW & 37 & 17& 0.56 & 1.29 & 1.03 \\ 
M2129 $\Omega_{\rm m},w_0, w_{\rm a}$-gNFW           & 36 & 18& 0.56 & 1.32 & 1.04 \\ 
\hline
\end{tabular}
\label{tab:cosmo_models_gnfw}
\tablefoot{Same as Table \ref{tab:cosmo_models_nfw} but for the gNFW model.}
\end{table}

\end{appendix}

\end{document}